\documentclass[11pt]{article}
\usepackage{lineno}
\usepackage{multirow}
\usepackage{amsmath}
\usepackage{float}
\usepackage{graphicx} 
\usepackage{amssymb,amscd,amsmath,amsthm,color}
\usepackage[T1]{fontenc} 
\usepackage{subcaption}
\usepackage{enumitem}
\usepackage{mathtools}
\usepackage{tcolorbox}
\usepackage{verbatim}
\usepackage{mathrsfs}
\usepackage{physics}
\usepackage{mathtools}
\usepackage{slashed}
\usepackage{longtable}
\usepackage{jheppub}
\usepackage{enumerate}
\usepackage{hyperref}
\usepackage{cleveref}
\newcommand{\Slash}[1]{\not\!#1}

\numberwithin{equation}{section}

\def\<{\langle}
\def\>{\rangle}

\def\be{\begin{equation}}
\def\ee{\end{equation}}

\def\cN{\mathcal{N}}

\def\cK{\mathcal{K}}

\newcommand{\ben}{\begin{eqnarray}\displaystyle}
\newcommand{\een}{\end{eqnarray}}

\makeatletter
\g@addto@macro\bfseries{\boldmath}\g@addto@macro\bfseries{\boldmath}
\makeatother
\preprint{CQUeST-2025-0765}
\title{\boldmath BPS Solutions of 4d Euclidean $\mathcal{N}=2$ Supergravity with Higher Derivative Interactions}

\author{Soumya Adhikari$^a$, Abhinava Bhattacharjee$^b$, and Amitabh Virmani$^c$}
\emailAdd{
ssoumya.a012@gmail.com, abhinava19@iisertvm.ac.in, avirmani@cmi.ac.in}
\affiliation[a]{Department of Physics \& Center for Quantum Spacetime, Sogang University\,,\\ 35 Baekbeom-ro, Mapo-gu, Seoul 04107, Republic of Korea}
\affiliation[b]{School of Physics, Indian Institute of Science Education and Research Thiruvananthapuram,\\
Thiruvananthapuram 695551 India}
\affiliation[c]{Chennai Mathematical Institute, 
H1, SIPCOT IT Park, Siruseri, Kelambakkam 603103 India}

\abstract{We study fully BPS and a broad class of half-BPS stationary  configurations of four-dimensional Euclidean $\mathcal{N}=2$ supergravity with higher-derivative interactions. Working within the off-shell conformal supergravity framework of de Wit and Reys (arXiv:1706.04973), we analyse the complete set of Killing spinor equations and obtain the corresponding algebraic and differential constraints. We further derive the Euclidean attractor equations and evaluate the Wald entropy for the fully BPS 
$AdS_{2}\times S^{2}$ background. For half-BPS stationary configurations, we obtain the generalized stabilization equations expressing all fields in terms of harmonic functions on three-dimensional flat base space, extending the Lorentzian analysis of  Cardoso et al.~(arXiv:hep-th/0009234) to the Euclidean signature. Our results provide a framework for studying supersymmetric saddles and computing the gravitational indices entirely within Euclidean higher-derivative supergravity, without recourse to analytic continuation.}

\makeatletter
\gdef\@fpheader{}
\makeatother
\begin{document}
\allowdisplaybreaks
\maketitle
\flushbottom
\section{Introduction}
\label{sec:intro} 
One of the major successes of string theory is the explanation of the entropy of a class of supersymmetric black holes as a count of microstates \cite{Sen:1995in, Strominger:1996sh, Sen:2007qy}.
This counting is performed by computing a supersymmetric index in a weakly coupled string theory. At strong coupling, the same system is described by supergravity, which admits supersymmetric black holes as classical solutions.

Underlying this remarkable matching, there is a conceptual tension in relating the entropy of the supersymmetric black hole to the index computed in the weakly coupled theory. Is the comparison justified? For a long time, we did not know how to compute the index on the gravity side. In recent years, this problem has been addressed~\cite{Cabo-Bizet:2018ehj, Iliesiu:2021are} by first defining the gravitational index in the Euclidean path integral approach to quantum gravity, and then identifying
the corresponding supersymmetric saddle geometries that contribute to the index of supersymmetric black holes. 

To set up the gravitational path integral that computes the index, we must not only make the time circle periodic but also set the Euclidean angular velocity to be such that fermions are periodic along the time circle. The corresponding saddle solutions preserve supersymmetry. Such saddles (often called index saddles) have been found in asymptotically AdS space and in asymptotically flat space \cite{Cabo-Bizet:2018ehj,Iliesiu:2021are,Cassani:2019mms,Bobev:2020pjk,Hristov:2022pmo, H:2023qko,Anupam:2023yns,Boruch:2023gfn,Hegde:2023jmp,Chowdhury:2024ngg,Chen:2024gmc,Cassani:2024kjn,Hegde:2024bmb,Adhikari:2024zif,Boruch:2025qdq,Bandyopadhyay:2025jbc,Boruch:2025biv, Cassani:2025iix, Boruch:2025sie}; for a review and further references, see \cite{Cassani:2025sim}. These saddle solutions are smooth, supersymmetric, yet at finite temperature, and perhaps most strikingly, they are typically complex in the Lorentzian signature. In some cases, these saddle solutions are real in the Euclidean theory, see, e.g., \cite{Hegde:2023jmp}.

In the context of $\mathcal{N}=2$ supergravity, the saddle solutions are half-BPS. These solutions have many remarkable properties. Two of these properties deserve special mention:  first, these solutions exhibit a new form of attraction \cite{Boruch:2023gfn, Chen:2024gmc}; second, in the higher derivative $\mathcal{N}=2$ supergravity, although we know almost nothing in detail about the saddle solutions, through a detailed analysis of the equations of motion it is possible to show that the gravitational index equals the Wald entropy \cite{Hegde:2024bmb}. The matching shown in \cite{Hegde:2024bmb} requires certain analytic continuation. A key open question is whether this matching can be understood intrinsically within the Euclidean theory, without invoking analytic continuation.

The goal of this paper is to take several steps toward answering this question. 
A more precise understanding of supersymmetric Euclidean configurations is also crucial for extending supersymmetric localization to gravitational theories. The success of counting of black hole microstates in string theory has been made more precise using supersymmetric localisation \cite{Dabholkar:2010uh,Sen:2008vm}. If we want to extend such results to understand supersymmetric localisation for computing the gravitational index, we need a better understanding of supersymmetric solutions in Euclidean supergravity.

 Motivated by these developments, in this work, we study equations of motion for fully BPS and a broad class of half-BPS stationary solutions of higher-derivative $\mathcal{N}=2$ Euclidean supergravity. Specifically,  we adapt the Lorentzian analysis of \cite{LopesCardoso:2000qm} to the Euclidean setting. The supergravity theories we consider are based on vector multiplets coupled to supergravity fields as constructed by de Wit and Reys \cite{deWit:2017cle}.  They constructed an off-shell Euclidean supergravity by carrying out an off-shell timelike reduction of five-dimensional off-shell Lorentzian supergravity. The construction is facilitated by the fact that both five-dimensional Lorentzian and four-dimensional Euclidean supergravities are based on symplectic Majorana spinors. A similar strategy is followed in several other works \cite{Cortes:2003zd, Cortes:2005uq, Cortes:2009cs}, but the reduction is typically performed on-shell.   As emphasized in \cite{Cortes:2003zd, Cortes:2005uq, Cortes:2009cs, deWit:2017cle, Jeon:2018kec, Ciceri:2023mjl}, in order to be compatible with supersymmetry, not only fermionic but also  bosonic fields in different spacetime signatures have different reality properties. These reality conditions make the analytic continuation rules from Lorentzian to Euclidean supergravity subtle \cite{Jeon:2018kec, Ciceri:2023mjl}.

Our methodology is as follows. We start with Euclidean conformal supergravity as constructed by de Wit and Reys.  We find the Killing spinor equations using superconformal transformations of the fermions. We then impose appropriate gauge fixing conditions to get to the equations satisfied by the solutions of the Euclidean theory. This approach has been used to classify stationary half-BPS solutions in $\cN=2$ higher derivative Poincar\'e supergravity \cite{LopesCardoso:2000qm, LopesCardoso:1998tkj, Mohaupt:2000mj}, and recently in finding fully supersymmetric solutions in higher derivative $\cN=4$ Poincar\'e supergravity~\cite{Bhattacharjee:2025qro}.

Needless to say, our analysis is fairly technical. A concise summary is as follows.  We obtain equations satisfied by fully BPS and a broad class of half-BPS stationary solutions. We derive the Euclidean attractor equations and evaluate the Wald entropy for Euclidean $AdS_{2}\times S^{2}$.  We derive the generalised stabilisation equations that express the half-BPS solutions in terms of harmonic functions. Our analysis forms the basis for future investigations. Using the results of this paper, we plan to explore the new attractor
mechanism in Euclidean supergravity with higher-derivative terms in our future work. We also plan to adapt the analysis of \cite{Hegde:2024bmb} to Euclidean
supergravity, and at the same time develop a detailed dictionary between Lorentzian and Euclidean variables, extending the analysis of \cite{Jeon:2018kec, Ciceri:2023mjl}.

The rest of the paper is organised as follows. In section \ref{N2sugra}, we give a brief review of Euclidean $\mathcal{N}=2$ supergravity in the framework of conformal supergravity and provide the relevant equations necessary for our analysis. In section \ref{sec:Full}, 
we assume the existence of two independent Killing spinors and analyse superconformal transformations of the fermions. We then gauge fix the extra symmetries of the off-shell 
theory and 
obtain fully supersymmetric configurations of the on-shell Euclidean $\mathcal{N}=2$ supergravity. In section \ref{sec:half}, we choose an embedding condition that makes one of the Killing spinors dependent on the other. We obtain constraints that describe such half-BPS configurations. In section \ref{sec:summary}, we collect the essential equations required to construct half-BPS solutions in terms of harmonic functions on the flat three-dimensional base space. In section \ref{conclusion}, we conclude the paper with a summary and some future directions. Our conventions are listed in appendix \ref{app:conventions}.

\section{\texorpdfstring{Euclidean $\cN=2$ supergravity}{Euclidean N=2 supergravity}}
\label{N2sugra}
 The construction of off-shell higher derivative Poincar\'e supergravity is facilitated by techniques of conformal supergravity. For constructing a version of $\cN=2$ Euclidean supergravity coupled with $n$ vector multiplets, we need a Euclidean Weyl multiplet along with $(n+1)$ Euclidean vector multiplets and a Euclidean hypermultiplet. One of the $(n+1)$ vector multiplets, together with the hypermultiplet, compensates for the extra gauge symmetries.

In this section, we briefly review 4d Euclidean $\cN=2$ conformal supergravity as developed in \cite{deWit:2017cle}. The construction in \cite{deWit:2017cle} proceeds via off-shell timelike reduction of five-dimensional $\cN=1$ Lorentzian supergravity to $\cN=2$ Euclidean supergravity in four dimensions. We present relevant multiplets of the theory, which include the $\cN=2$ Weyl multiplet, $\cN=2$ vector multiplets, and $\cN=2$ hypermultiplets, focusing on the transformation rules. 

The Weyl multiplet \cite{Ferrara:1977mv, deWit:1979dzm, Bergshoeff:1980is, Bergshoeff:1985mz,Fujita:2001kv, Butter:2013goa, Butter:2013rba} is the most crucial multiplet for conformal supergravity. It contains the graviton, its superpartner, the gravitino and other gauge fields along with a set of auxiliary fields. For the Euclidean $\cN=2$ conformal supergravity in four dimensions, the Weyl multiplet contains superconformal gauge fields corresponding to the superconformal group SU$^*(4|2)$. Independent gauge fields are the vielbein $e_{\mu}{}^{a}$, the dilatation gauge field $b_\mu$, a chiral SU(2) gauge field $\mathcal{V}_{\mu}{}^{i}{}_{j}$, a chiral SO$(1,1)$ gauge field $A_{\mu}$, and the $Q$-supersymmetry gauge field $\psi_{\mu}{}^{i}$. The composite gauge fields are the spin connection $\omega_{\mu}{}^{ab}$, the gauge field associated with the special conformal transformation $f_{\mu}{}^a$, and the $S$-supersymmetry gauge field $\phi_{\mu}{}^{i}$. In addition, the Weyl multiplet contains auxiliary matter fields: an antisymmetric tensor $T_{ab}$, a symplectic Majorana spinor doublet $\chi^i$, and a scalar $D$.  Here $i,j=1,2$ are the SU(2) indices, $a,b=0,\cdots,3$ are the tangent space indices, and $\mu$ is a Euclidean world index. 

In the Euclidean theory, the spinors are symplectic Majorana in nature. The symplectic Majorana condition is defined as follows. First, for Euclidean spinors the
Dirac conjugate is taken to be the Hermitian conjugation, so that $\bar \chi_i \equiv
\chi_i^\dagger$. With this definition, the symplectic Majorana spinors
satisfy the condition
\begin{equation}
  \label{eq:Majorana-4D}
  C^{-1} \,\bar\chi_i {}^{\rm T}= \varepsilon_{ij}\,\chi^j.
\end{equation}
The charge conjugation matrix $C$ is anti-symmetric and unitary. The (hermitian) gamma matrices $\gamma_a$ satisfy  the relation 
\begin{equation}
  \label{eq:gamma-C}
  C\,\gamma_a\,C^{-1} = -\gamma_a{\!}^\mathrm{T}\,, \qquad
  (a=0, 1,2,3). 
\end{equation}
We define $\gamma_5\equiv \gamma_0
\,\gamma_1\,\gamma_2\,\gamma_3$ which satisfies the relation\footnote{There is one small difference between our notation and ref.~\cite{deWit:2017cle}. There $a=1,2,3,4$ whereas for us $a = 0,1,2,3$.} 
\begin{equation}
  \label{eq:gamma-5}
  C\,\gamma_5\,C^{-1} = \gamma_5{\!}^\mathrm{T}\,. 
\end{equation}

The fermions can be split into their chiral and anti-chiral parts. They are denoted by $+$ and $-$ in the subscript respectively. The independent gauge fields together with the auxiliary fields constitute 24 bosonic + 24 fermionic degrees of freedom. The details of the Weyl multiplet are given in Table \ref{table:weyl}. We denote the Weyl weight and the SO(1,1) chiral weight by $w$ and $c$ respectively.
\begin{table}[H]
\centering
\begin{tabular*}{14.8cm}{@{\extracolsep{\fill}}
    |c|cccccccc|ccc|ccc| }
\hline
 & &\multicolumn{9}{c}{Weyl multiplet} & &
 \multicolumn{2}{c}{parameters} & \\[1mm]  \hline \hline
 field & $e_\mu{}^{a}$ & $\psi_\mu{\!}^i$ & $b_\mu$ & $A_\mu$ &
 $\mathcal{V}_\mu{}^i{}_j$ & $T_{ab}^\pm $ &
 $ \chi^i $ & $D$ & $\omega_\mu^{\,ab}$ & $f_\mu{}^a$ & $\phi_\mu{\!}^i$ &
 $\epsilon^i$ & $\eta^i$
 & \\[.5mm] \hline
$w$  & $-1$ & $-\tfrac12 $ & 0 &  0 & 0 & 1 & $\tfrac{3}{2}$ & 2 & 0 &
1 & $\tfrac12 $ & $ -\tfrac12 $  & $ \tfrac12  $ & \\[.5mm] \hline
$c$  & $0$ & $\mp\tfrac12 $ & 0 &  0 & 0 & $\pm1$ & $\mp\tfrac{1}{2}$ & 0 &
0 & 0 & $\pm\tfrac12 $ & $ \mp\tfrac12 $  & $ \pm\tfrac12  $ & \\[.5mm] \hline
 $\gamma_5$   &  & $\pm$ &   &    &   &  & $\pm$ &  & &
 & $\pm$ & $\pm$  & $\pm $ & \\ \hline
\end{tabular*}
\vskip 2mm
\renewcommand{\baselinestretch}{1}
\parbox[c]{14.8cm}{\caption{\label{table:weyl}{
  $\cN=2$ Euclidean Weyl multiplet in four dimensions.  The Weyl weights and the SO(1,1) chiral weights are listed. The $\gamma_5$ chiralities  of the spinors are also listed. }}}
\end{table}

The complete supersymmetry transformations of the Weyl multiplet are given in \cite{deWit:2017cle}. For our purposes, we only need the supersymmetry transformations of fermions which are given below. The $Q$ and $S$-supersymmetry transformations are parametrized by symplectic Majorana spinors $\epsilon^i$ and $\eta^{i}$, respectively. We have,
\begin{align}\label{weylvari}
     \delta\psi_\mu{}^i =&\,2\,\mathcal{D}_\mu \epsilon^i  
    + \frac1{16} \mathrm{i}  \,(T_{ab}{\!}^+ +T_{ab}{\!}^- ) \gamma^{ab} \gamma_\mu
    \epsilon^i    -\mathrm{i} \gamma_\mu \eta^i, \\
    \delta \chi^{i}=\,&\,  \frac1{24}
  \mathrm{i} \,\gamma^{ab}\Slash{D}
  (T_{ab}^+ + T_{ab}^-)  \epsilon^i+
  \frac1{6}R(\mathcal{V})_{ab}{\!}^i{}_j \,\gamma^{ab} \epsilon^j
  -\frac13 R(A)_{ab} \,\gamma^{ab} \gamma^5 \epsilon^i \nonumber\\ 
  &\,  + D\,
  \epsilon^i +\frac1{24}(T_{ab}^+ + T_{ab}^-) \gamma^{ab} \eta^i .
\end{align}
The derivative $D_{\mu}$ appearing here and elsewhere is the fully superconformal covariant derivative and the derivative $\mathcal{D}_{\mu}$ is covariant with respect to all bosonic symmetries except the special conformal transformation. $R(\mathcal{V})_{ab}{\!}^i{}_j$ and $R(A)_{ab}$ are the  supercovariant curvature tensors for the gauge fields $\mathcal{V}_{\mu}{}^{i}{}_{j}$ and $A_{\mu}$ respectively. 

The supercovariant curvature tensors associated with $e_{\mu}{}^{a}, \omega_{\mu}{}^{ab}$ and $\psi_{\mu}{}^{i}$ are,
\begin{align}
     R(P)_{\mu \nu}{}^a =\,&\; 2 \, \mathcal{D}_{[\mu} \, e_{\nu]}{}^a
  - \frac1{2}\,\bar{\psi}_{i[\mu}\gamma^5\gamma^a\psi_{\nu]}{}^i,\\
   R(Q)_{\mu \nu}{}^i =\,& \; 2 \, \mathcal{D}_{[\mu} \psi_{\nu]}{}^i -
  \mathrm{i}\,\gamma_{[\mu} \phi_{\nu]}{}^i +
  \frac{1}{16}\mathrm{i}\,(T_{ab}^+ + T_{ab}^-) \,
  \gamma^{ab}\gamma_{[\mu} \psi_{\nu]}{}^i,\\
   R(M)_{\mu \nu}{\!}^{ab} =& \; 2 \,\partial_{[\mu} \omega_{\nu]}{}^{ab}
  - 2\, \omega_{[\mu}{}^{ac}\omega_{\nu]c}{}^b - 4 f_{[\mu}{}^{[a}
  e_{\nu]}{}^{b]} +
\frac12\mathrm{i}\,\bar{\psi}_{i[\mu}\gamma^5\phi_{\nu]}{}^i\nonumber\\
  &\,
  -\frac18\mathrm{i}\,\bar{\psi}_{\mu\,i}\gamma^5\psi_\nu{}^i\,(T^{ab+}
  + T^{ab-}) -
\frac34\bar{\psi}_{i[\mu}\gamma^5\gamma_{\nu]}\gamma^{ab}\chi^i -
\bar{\psi}_{i[\mu}\gamma^5\gamma_{\nu]}R(Q)^{ab\,i}.
\end{align}
The definitions of the remaining supercovariant curvature tensors can be found in \cite{deWit:2017cle}.  To get a theory of gravity from the superconformal gauge theory, we need to impose three conventional constraints on the supercovariant curvatures,
\begin{align}
      &R(P)_{\mu \nu}{}^a =  0 \, ,  \\[1mm]
  &\gamma^\mu R(Q)_{\mu \nu}{}^i + \frac32 \gamma_{\nu}
  \chi^i = 0 \, , \\[1mm]
  &
  e^{\nu}{\!}_b \,R(M)_{\mu \nu}{\!}^{ab} - \widetilde{R}(A)_{\mu}{\!}^a 
  + \frac1{16} T_{\mu{b}}{\!}^-  \,T^{ab+}  
  -\frac{3}{2}\, D \, e_\mu{\!}^a = 0 \, .
\end{align} 
We use the notation where $\widetilde R_{ab}$ equals the dual tensor $\frac12 \varepsilon_{abcd}R^{cd}$. The above constraints make the gauge fields $\omega_{\mu}{}^{ab}, f_{\mu}{}^{a}$ and $\phi_{\mu}{}^{i}$ dependent. 

Expressions for the dependent gauge fields are given in \cite{deWit:2017cle}. We only need the expression for $f_{\mu}{}^{a}$ and $\omega_{\mu}{}^{ab}$
    \begin{align}
     f_\mu{\!}^{a} =\,& \frac12\,R(\omega,e)_\mu{}^a -
  \frac14\,\left(D+\frac13 R(\omega,e)\right) e_\mu{}^a -
  \frac12\,\widetilde{R}(A)_\mu{}^a - \frac1{32}\,T_{\mu
    b}^-\,T^{+\,ba} + \cdots,\label{fmua}\\
     \omega_\mu{\!}^{ab} =\,& -2\,e^{\nu[a} \,(\partial_{[\mu} + b_{[\mu}) e_{\nu]}{\!}^{b]}
     -e^{\nu[a}e^{b]\rho}\,e_{\mu c}\,(\partial_\rho +b_\rho) e_\nu{\!}^c +\cdots
\end{align}
 where the ellipsis contains fermionic contributions. The uncontracted curvature $R(\omega)_{\mu \nu}{}^{ab}$ is defined as
 \begin{equation}\label{riccicurvature}
R(\omega)_{\mu \nu}{}^{ab} = \partial_\mu \omega_\nu{}^{ab} - \partial_\nu \omega_\mu{}^{ab} -  \omega_\mu{}^{ac} \omega_{\nu c}{}^{b} + \omega_\nu{}^{ac} \omega_{\mu c}{}^{b}.
 \end{equation} Moreover, $R(\omega,e)_{\mu}{}^{a}=R(\omega)_{\mu \nu}{}^{ab} e_{b}{}^{\nu}$ is the non-symmetric Ricci tensor, and $R(\omega, e)$, the corresponding Ricci scalar. The following modified curvature features prominently in the following:
\begin{align}\label{curlyR}
    \mathcal{R}(M)_{ab}{}^{cd}=R(M)_{ab}{}^{cd}+\frac{1}{32}\left(T_{ab}^{-}T^{+cd}+T_{ab}^{+}T^{-cd}\right).
\end{align}

 The  $Q$- and $S$-supersymmetry variation of the gravitino field strength is given as,
\begin{align}
    \delta R(Q)_{ab\pm}{}^i=\frac{1}{4} \mathrm{i} \slashed{D}T_{ab}^{\mp}\epsilon_{\mp}^{i}+R(\mathcal{V})_{ab}^{\mp\,i}{}_{j}\epsilon_{\pm}^j-\frac{1}{2}\mathcal{R}(M)_{ab}{}^{cd}\gamma_{cd}\epsilon_{\pm}^{i}
    +\frac{1}{16}T_{cd}{}^{\mp}\gamma^{cd}\gamma_{ab}\eta_{\pm}^{i}.
\end{align}
The $Q$- and $S$-supersymmetry variation of the composite gauge field $\phi_{\mu}{}^i$ is also needed for the analysis below. It takes the form,
 \begin{align}
       \delta\phi_\mu{}^i =&\, 2\,\mathcal{D}_\mu \eta^i + 2\mathrm{i}
  \,f_\mu{}^a\gamma_a\epsilon^i 
  +\frac1{16}\Slash{D}  (T_{ab}{\!}^++T_{ab}{\!}^-) \gamma^{ab}\gamma_\mu 
 \,\epsilon^i \nonumber\\
  &\, -\frac14\mathrm{i}\gamma^{ab}\gamma_\mu
  R(\mathcal{V})_{ab}{\!}^i{}_j \,\epsilon^j
  - \frac12\mathrm{i}\gamma^{ab}\gamma_\mu \gamma^5 
  R(A)_{ab}\,\epsilon^i \nonumber\\
  &\,
  -\frac32 \mathrm{i} [ ( \bar{\chi}_j \gamma^5 \gamma^a \epsilon^j )
  \gamma_a \psi_\mu{}^i - ( \bar{\chi}_j \gamma^5 \gamma^a
  \psi_\mu{}^j ) \gamma_a \epsilon^i ]. 
 \end{align}

A chiral (anti-chiral) multiplet constitutes 8 + 8 bosonic and fermionic degrees of freedom. The components are denoted as (plus for chiral, minus for anti-chiral)
 \begin{align}
     \left(A_{\pm}, \Psi_{\pm}^{i}, B_{\pm}^{ij}, F_{ab}^{\mp}, \Lambda^{i}_{\pm}, C_{\pm}\right).
 \end{align}
Relevant $Q$- and $S$-supersymmetry transformations are given below. The rest of the transformations can be found in \cite{deWit:2017cle}. We have,
 \begin{align}
    \delta A_{\pm}=\,&\pm \mathrm{i} \Bar{\epsilon}_{i\pm}\Psi^{i}_{\pm}\label{Atrans}, \\
    \delta \Psi_{\pm}^{i}=\,&-2\mathrm{i} \slashed{D}A_{\pm}\epsilon_{\mp}^{i}+\varepsilon_{jk}B^{ij}{}_{\pm}\epsilon^{k}{}_{\pm}-\frac{1}{2}F_{ab}{}^{\mp}\gamma^{ab}\epsilon^i{}_{\pm}+2 w A_{\pm}\eta_{\pm}^{i}, \\ 
    \delta \Lambda^i{}_{\pm}=\,& \frac{1}{2}\mathrm{i} \gamma^{ab}\slashed{D}F_{ab}^{\mp}\epsilon^i_{\mp}+ \mathrm{i}  \slashed{D}B^{ij}{}_{\pm}\varepsilon_{jk}\epsilon^{k}_{\mp}+ C_{\pm}\epsilon^{i}_{\pm} \nonumber\\
    \,& +\frac{1}{8}\mathrm{i} (\slashed{D}A_{\pm}T_{ab}^{\pm}+ w A_{\pm}\slashed{D} T_{ab}^{\pm})\gamma^{ab}\epsilon^{i}_{\mp}\pm \frac{3}{2}\mathrm{i}  \gamma_a\epsilon^i{}_{\mp}\Bar{\chi}_{j\mp}\gamma^a\Psi^j{}_{\pm}\nonumber \\
    \,& -(1+w)B^{ij}{}_{\pm}\varepsilon_{jk}\eta^{k}_{\pm}+\frac{1}{2}(1-w)\gamma^{ab}F_{ab}^{\mp}\eta_{\pm}^{i}.
    \end{align}
Here $w=-c$ is the Weyl weight of the lowest component $A_{+}$ of the chiral multiplet; likewise $w=c$ for the anti-chiral multiplet field $A_-$.

The Weyl multiplet is a combination of a chiral anti-self-dual tensor multiplet and an anti-chiral self-dual tensor multiplet. The self-dual and anti-self-dual multiplets both have $24 + 24$ degrees of freedom. Imposing a constraint on this reducible field representation results in the Weyl multiplet with $24 + 24$ degrees of freedom. Considering the square of the Weyl multiplet, one can construct a scalar chiral and anti-chiral multiplet of
weight $w = 2$. The components are as follows,
\begin{align}
  \label{eq:weyl-squared-components}
  A_\pm =&\, \big(T_{ab}{\!}^\mp\big)^2 \, ,  \\[1mm]
  \Psi^i{\!}_\pm =&\, -16\,T^{\mp\,ab}\,R(Q)_{ab}{\!}^i{\!}_\pm
  \, ,  \\[1mm] 
  B^{ij}{\!}_\pm=&\, 16\,T^{\mp\,ab} \varepsilon^{k(i} \,
  R(\mathcal{V})_{ab}{}^{j)}{}_k \mp
  64\mathrm{i}\,\varepsilon^{k(i}\, \bar{R}(Q)_{ab\,k\,\pm}\,R(Q)^{ab\,j)}{\!}_\pm
  \, ,  \\[1mm]
  F_{ab}{\!}^\mp=&\,
  -16\,\mathcal{R}(M)^{cd}{\!}_{ab}\,T _{cd}{\!}^{\mp} 
  \pm
  16\mathrm{i}\,\bar{R}(Q)_{cd\,i\,\pm}\,\gamma_{ab}\,R(Q)^{cd\,i}{\!}_\pm
  \, , \\[1mm]
  \Lambda^i{\!}_\pm =&\, -32\,\gamma^{ab} R(Q)_{cd}{}^i{}_\pm
 \mathcal{R}(M)^{cd}{}_{ab} + 16\,\big(\mathcal{R}(S)_{ab}{}^i{}_\pm - 3\mathrm{i}\,\gamma_{[a}D_{b]}\,\chi^i_\mp
 \big)T^{\mp ab}  \nonumber\\
 &\, - 64\,R(\mathcal{V})_{ab}{}^i{}_j\,R(Q)^{ab\,j}{\!}_\pm \, , \\[1mm]
  C_\pm =&\,
  -64\,\mathcal{R}(M)^{cd\mp}{\!}_{ab}\,\mathcal{R}(M)_{cd}{\!}^{\mp \,ab}
  - 32\,R(\mathcal{V})^{\mp\,ab\,i}{}_j
  \,R(\mathcal{V})_{ab}{\!}^{\mp j}{}_i 
  \nonumber\\
   &\, +16\, T^{\mp\,ab}D_a D^c T_{cb}{\!}^\pm 
    \mp
    128\mathrm{i}\,\bar{\mathcal{R}}(S)^{ab}{}_{i\pm}\,R(Q)_{ab}{}^i{}_\pm
    \mp 384\,\bar{R}(Q)^{ab}{}_{i\pm}\gamma_a D_b\chi^i{\!}_\mp \, .  
\end{align}

Euclidean vector multiplets are labelled by an index $I=0,\ldots, n$. A vector multiplet
 contains two real scalar fields $X^I_{+}$ and $X^I_{-}$, one symplectic Majorana spinor $\Omega^{Ii}$ (gauginos), a U(1) gauge field $W_{\mu}^I$, and an auxiliary field $Y^{ij\,I}$ satisfying the pseudo-reality condition $(Y^{ij\,I})^* \equiv Y_{ij}{}^{I} =
\varepsilon_{ik}\varepsilon_{jl}\,Y^{kl\,I}$. The $Q$- and $S$-supersymmetry transformations of  $\Omega_{\pm}^{Ii}$ and $X_{\pm}^{I}$ are given as,
\begin{align}
    \delta \Omega^{Ii}_{\pm}=\,& -2 \mathrm{i} \, \slashed{D}X_{\pm}^{I}\epsilon^{i}_{\mp}-\frac{1}{2}\left[{F}^{I}_{ab}{}^{\mp}-\frac{1}{4}X^{I}_{\mp}T_{ab}{}^{\mp}\right]\gamma^{ab}\epsilon_{\pm}^{i}-\varepsilon_{kj}Y^{ ik\,I}\epsilon^j{}_{\pm} + 2 X^{I}_{\pm}\eta^i_{\pm},\\
     \delta X_{\pm}^I=\,&\pm \mathrm{i} \, \Bar{\epsilon}_{i\pm}\Omega_{\pm}^{Ii }\label{Xpm}.
\end{align}
where 
the supercovariant field strength ${F}^{I}_{\mu\nu}$ corresponding to the U(1) gauge field $W^{I}_{\mu}$ is defined as
\begin{align}
{F}^{I}_{\mu\nu}=&2\partial_{[\mu}W^{I}_{\nu]}+ \Bar{\psi}_{i[\mu}\gamma_{\nu]}\Omega^{Ii}_{+}-\Bar{\psi}_{i[\mu}\gamma_{\nu]}\Omega^{Ii}_{-} +\mathrm{i} \, X^{I}_{-}\Bar{\psi}_{\mu i}{\psi_{\nu}}^i_{+}- \mathrm{i} \, X^{I}_{+}\Bar{\psi}_{\mu i}{\psi_{\nu}}^i_{-}.
\end{align} 

 A Euclidean hypermultiplet contains two scalars and a symplectic Majorana spinor. The fields for $r$  Euclidean hypermultiplets are arranged as follows.
Let $\alpha = 1, \ldots, 2r$. The bosonic fields are $A_{i}{}^{\alpha}$ (recall $i=1,2$) and the spinor fields are $\zeta^{\alpha}$. The bosonic fields are subjected to the pseudo-reality constraint,
\begin{align}
    (A_{j}{}^{\beta})^{*}=A^{j}{}_{\beta}=A_{i}{}^{\alpha}\varepsilon^{ij}\Omega_{\alpha\beta},
\end{align}
where $\Omega_{\alpha \beta}$ is an antisymmetric matrix. The spinor fields $\zeta^{\alpha}$ are subjected to the symplectic Majorana condition with respect to the matrix  $\Omega_{\alpha \beta}$. The spinor fields  transform under $Q$- and $S$- supersymmetry as follows,
\begin{align}\label{transzeta}
    \delta \zeta^{\alpha}_{\pm}=\,&- \mathrm{i} \slashed{D}A_{i}{}^{\alpha}\epsilon^i_{\mp}+ A_{i}{}^{\alpha}\eta^{i}_{\pm}.
\end{align}
For our later analysis, we need the supercovariant derivative of the spinors $\zeta_{\pm}^{\alpha}$. It is given as, 
\begin{align}\label{zetaalpha}
D_{\mu}\zeta^\alpha_{\pm}=\,&\mathcal{D}_{\mu}\zeta^{\alpha}_{\pm}+\frac{1}{2}\mathrm{i}\slashed{\mathcal{D}}A_{i}{}^{\alpha}\psi_{\mu\mp}^i-\frac{1}{2}A_{i}{}^{\alpha}\phi_{\mu\pm}^{i}.
\end{align}

 It is also convenient to introduce the hyper-K\"ahler potential $\chi$,
\begin{equation}
    \varepsilon_{ij} \chi = {\Omega}_{\alpha \beta} A_i^\alpha A_j^\beta.
\end{equation}

We now write down the supergravity Lagrangian \cite{deWit:2017cle}.  Unlike the Lorentzian case, two independent real prepotential functions $\mathcal{F}^{\pm}(X^I_{\pm},A_{\pm})$ appear in the Euclidean case. These functions are homogeneous of degree two, 
\begin{align}
    \mathcal{F}^{\pm}(\lambda X_{\pm}^{I}, \lambda^{2}A_{\pm})=\lambda^2\mathcal{F}^{\pm}(X_{\pm}^{I},A_{\pm}),
\end{align}
and they follow the homogeneity property
\begin{align}\label{homo}
\mathcal{F}_{I}^{\pm}=X^{J}_{\pm}\mathcal{F}_{IJ}^{\pm}+2 A_{\pm}\mathcal{F}^{\pm}_{IA}.
\end{align}
The scalars of the vector multiplet parametrize a para-K\"ahler manifold \cite{Cortes:2003zd} with K\"ahler potential given as 
\begin{equation}\label{kahlar}
    e^{-\mathcal{K}}=\mathcal{F}^{+}_{I}(X_{+},A_{+})X_{-}^{I}+X_{+}^{I}\mathcal{F}^{-}_{I}(X_{-}, A_{-}).
\end{equation}
The Lagrangian reads
\begin{align}\label{action}
    \mathcal{L}=\mathcal{L}_{+}+\mathcal{L}_{-}+\mathcal{L}_\mathrm{H}+\mathcal{L}_{A+}+\mathcal{L}_{A-},
\end{align}
where the vector multiplet Lagrangian terms are $\mathcal{L}_\pm$
\begin{align}
  \label{eq:vector-lagrangian-pm}
  e^{-1} \mathcal{L}_\pm = &\, -\mathcal{F}^\pm_I \,D_{c}D^{c}\,
  X_\mp^I - \frac18 \mathcal{F}^\pm_I\,\left(F_{ab}^{
    I\pm} - \frac14\,X_\pm^I T_{ab}{\!}^\pm\right)\,T^{ab\,\pm}
  \nonumber\\[1mm]
  &\, +\frac14\,\mathcal{F}^\pm_{IJ}\, \left(F_{ab}^{I\mp} -
  \frac14\,X_\mp^I\, T_{ab}{\!}^\mp\right) \,\left(F^{ab\,J\mp}
  - \frac14\,X_\mp{\!}^{J}\, T^{ab\,\mp}\right) \nonumber \\
  &\, +\frac18\,\mathcal{F}^\pm_{IJ}\,Y^{ij\,I}\, Y_{ij}{\!}^J
  -\frac1{32}\,\mathcal{F}^\pm\, (T_{ab}{\!}^\pm)^2 \,.
\end{align}
The chiral (anti-chiral) multiplet Lagrangian terms are,
\begin{align}
  \label{eq:chiral-bckgd-Lagrangian-pm}
  e^{-1}\mathcal{L}_{A\pm} =&\, -\frac12\,
  \mathcal{F}^\pm_{{A}}\,{C}_\pm -
  \frac14\,\mathcal{F}^\pm_{{A}I}\,{B}_\pm{\!}^{ij}
  \,Y_{ij}{\!}^I 
  +\frac12\,\mathcal{F}^\pm_{{A}I}\,
  {F}^{\,ab\,\mp}\left(F_{ab}^{I \mp} 
  - \frac14\,X_\mp^I
  T_{ab}{\!}^\mp\right) \nonumber \\[1mm]
  &\,
  +\frac18\,\mathcal{F}^\pm_{AA}\,
  {B}_\pm{\!}^{ij}\,{B}_\pm{\!}^{kl}
  \,\varepsilon_{ik}\,\varepsilon_{jl} +
  \frac14\,\mathcal{F}^\pm_{AA}\,{F}_{ab}{\!}^\mp
  \,{F}^{\,ab\,\mp} \, ,
\end{align} 
and the hypermultiplet Lagrangian term is
\begin{align}
  \label{eq:hyper-lagrangian}
  e^{-1}\mathcal{L}_\mathrm{H} =&\;
  \frac12\,\varepsilon^{ij}\,\Omega_{\alpha\beta}\,A_i{}^\alpha \left(D^a
  D_a + \frac32\,D\right)
  A_j{}^\beta .
\end{align}
The second supercovariant derivatives of the hypermultiplet scalars and vector multiplet scalars 
that appear in \eqref{eq:vector-lagrangian-pm} and \eqref{eq:hyper-lagrangian}, are  defined as,
\begin{align}
    D_{a}D^{b}A_{i}{}^{\alpha}=\,&\mathcal{D}_{a}\mathcal{D}^{b}A_{i}{}^{\alpha}+f_{a}{}^{b}A_{i}^{\alpha},\\[1mm]
    D_{a}D^{b}X_{\pm}^{I}=\,&\mathcal{D}_{a}\mathcal{D}^{b}X_{\pm}^{I}+f_{a}{}^{b}X_{\pm}^{I}.
\end{align}
We now introduce dual electromagnetic field tensors as
\begin{align}\label{dual}
    G^{\mu\nu \pm}{}_{I}\,&\equiv\pm \frac{2}{e}\frac{\partial \mathcal{L}}{\partial F_{\mu\nu}^{I \pm }}\nonumber \\
    \,&=\pm\left[\mathcal{F}_{IJ}^{\mp} \left(F^{J \mu\nu \pm }-\frac{1}{4}X_{\pm}^{J}T^{\mu\nu \pm}\right)+\mathcal{F}_{IA}^{\mp}{F}^{\mu\nu\pm}-\frac{1}{4}\mathcal{F}_{I}^{\pm}T^{\mu\nu \pm}\right].
\end{align}
With these definitions, the equations of motion and Bianchi identities for the gauge fields take the form,
\begin{align}
    \mathcal{D}_{a}(G^{ab+}{}_{I}-G^{ab-}{}_{I})=\,&0\label{eom},\\
    \mathcal{D}_{a}(F^{abI+}-F^{abI-})=\,&0.\label{bian}
\end{align}
For later use, we note the following identities obtained by using \eqref{dual}, 
\begin{align}
    e^{\mathcal{K}}(\mathcal{F}_{I}^{+}F_{\mu\nu}^{I-}+X_{+}^{I}G^{-}_{\mu\nu I})-\frac{1}{4}T^{-}_{\mu\nu}=\,&e^{\mathcal{K}}\mathcal{F}_{IA}^{+}\left[2{A}_{+}\left(F_{\mu\nu}^{I-}-\frac{1}{4}X_{-}^{I}T_{\mu\nu}^{-}\right)-X_{+}^{I}{F}_{\mu\nu}^{-}\right],\label{iden}\\
     e^{\mathcal{K}}(\mathcal{F}_{I}^{-}F_{\mu\nu}^{I+}-X_{-}^{I}G^{+}_{\mu\nu I})-\frac{1}{4}T^{+}_{\mu\nu}=\,&e^{\mathcal{K}}\mathcal{F}_{IA}^{-}\left[2 {A}_{-}\left(F_{\mu\nu}^{I+}-\frac{1}{4}X_{+}^{I}T_{\mu\nu}^{+}\right)-X_{-}^{I}{F}_{\mu\nu}^{+}\right].\label{iden2}
\end{align}
Furthermore, we need the $D$ and the chiral U(1) gauge field $A_{\mu}$ equations of motion for the supersymmetry analysis of the following sections. These equations read,
\begin{align}
3e^{-\mathcal{K}}+\frac12 \chi=\,&-192D(\mathcal{F}_{A}^{+}+\mathcal{F}_{A}^{-})+4~ (T_{cd}^{-})^{-2}T^{ab-}\Big(\mathcal{F}_{I}^{+}F_{ab}^{I-}+X_{+}^{I}G_{ab I}^{-}\Big)\nonumber\\
\,&+4~ (T_{cd}^{+})^{-2}T^{ab+}\Big(\mathcal{F}_{I}^{-}F_{ab}^{I+}-X_{-}^{I}G_{ab I}^{+}\Big),\label{Deom}
\end{align}
\begin{align}
-e^{-\mathcal{K}}\mathcal{A}_{c}& ~ =~ -4\mathcal{D}_{a}(\mathcal{F}_{A}^{+}-\mathcal{F}_{A}^{-})T^{ab-}T_{cb}^{+}+4(\mathcal{F}_{A}^{+}+\mathcal{F}_{A}^{-})\Big(T_{cb}^{-}\mathcal{D}_{a}T^{ab+}-T_{cb}^{+}\mathcal{D}_{a}T^{ab-}\Big)\nonumber\\
\,&+8\mathcal{D}^{a}\Bigg[(T_{ef}^{-})^{-2}T^{b-}{}_{[a}\Big(\mathcal{F}_{I}^{+}F_{c]b}^{I-}+X_{+}^{I}G_{c]bI}^{-}\Big)-(T_{ef}^{+})^{-2}T^{b+}{}_{[a}\Big(\mathcal{F}_{I}^{-}F_{c]b}^{I+}-X_{-}^{I}G_{c]bI}^{+}\Big)\Bigg]\nonumber\\
\,&
+128 \mathcal{D}^{a}\big[\mathcal{F}_{A}^{+}R(A)^{-}_{ca}+\mathcal{F}_{A}^{-}R(A)^{+}_{ca}\big],\label{EOMAMU}
\end{align}
where we have defined 
\begin{align}
        \mathcal{A}_{\mu}\,&\equiv \frac{1}{2}e^{\mathcal{K}}\Big[\mathcal{F}_{I}^{+} \overset{\leftrightarrow}{\mathcal{D}}_{\mu}X_{-}^{I}+X_{+}^{I}\overset{\leftrightarrow}{\mathcal{D}}_{\mu}\mathcal{F}^{-}_{I}\Big].
\end{align}

\section{\texorpdfstring{Fully BPS solutions}{Fully-BPS solutions}}\label{sec:Full}
In this section, we analyse fully supersymmetric solutions of Euclidean $\mathcal{N}=2$ supergravity, following \cite{LopesCardoso:2000qm}. We analyse the Killing spinor equations, which we get by setting the $Q$-supersymmetry transformation of all independent spinors in the Euclidean theory to zero. A clever way to achieve this is to use the conformal supergravity set-up. Since the Euclidean supergravity comes from the corresponding conformal supergravity by appropriate gauge fixing, one can see that the $Q$-supersymmetry transformation rule in the Euclidean theory is a linear combination of $Q$- and $S$-supersymmetry transformation of the conformal theory.  We further note that setting $Q$-supersymmetry transformation in the Euclidean
theory to zero is equivalent to setting the conformal $Q$-supersymmetry transformation of $S$-supersymmetry invariant spinors to zero. Thus, we proceed by explicitly constructing $S$-invariant linear combinations and demand their $Q$-variations to vanish.

Typically, an $S$-invariant spinor can be obtained by taking an appropriate linear combination of a spinor with its $S$-supersymmetry compensating spinor. To construct our first $S$-supersymmetry compensator, we proceed as in \cite{LopesCardoso:2000qm}. We observe that the supersymmetric variation of the K\"ahler potential \eqref{kahlar} leads to a spinor, which can be used as a $S$-supersymmetry compensator. The $Q$-supersymmetry transformation of the K\"ahler potential $\mathcal{K}$ is,
\begin{align}
    \delta \mathcal{K}=-e^{\mathcal{K}}(\mathcal{F}^{+}_{I}\delta X_{-}^{I}+\mathcal{F}_{I}^{-}\delta X_{+}^{I}+\delta\mathcal{F}^{+}_{I} X_{-}^{I}+\delta\mathcal{F}_{I}^{-} X_{+}^{I}),
\end{align}
where using the transformation rules given in \eqref{Atrans} and \eqref{Xpm}, we obtain 
\begin{align}
    \delta \mathcal{K}=\Bar{\epsilon}_{i+}\zeta^{Vi}_{+}+\Bar{\epsilon}_{i-}\zeta^{Vi}_{-},
\end{align}
with the spinors $\zeta^{Vi}_{\pm}$ are defined as,
\begin{align}
    \zeta_{+}^{V i}\equiv\,&- \mathrm{i} e^{\mathcal{K}}\Big[(\mathcal{F}_{I}^{-}+X_{-}^{J}\mathcal{F}_{IJ}^{+})\Omega_{+}^{Ii}+X_{-}^{I}\mathcal{F}_{IA}^{+}\Psi_{+}^i\Big],\\
    \zeta_{-}^{V i}\equiv\,&\mathrm{i} e^{\mathcal{K}}\Big[(\mathcal{F}_{I}^{+}+X_{+}^{J}\mathcal{F}_{IJ}^{-})\Omega_{-}^{Ii}+X_{+}^{I}\mathcal{F}_{IA}^{-}\Psi_{-}^i\Big].
\end{align}
The $Q$- and $S$-supersymmetry transformations of $\zeta^{Vi}_{\pm}$ spinors are,
    \begin{align}
\delta\zeta_{+}^{V i}=\,&-e^{\mathcal{K}}\slashed{D}e^{-\mathcal{K}}\epsilon_{-}^{i}+2\slashed{\mathcal{A}}\epsilon_{-}^{i}+ {\frac{1}{2}\mathrm{i}\mathcal{F}_{ab}^{-}\gamma^{ab}\epsilon_{+}^{i}}\nonumber\\
\,&+\mathrm{i}e^{\mathcal{K}}\Big[(\mathcal{F}_{I}^{-}+X_{-}^{J}\mathcal{F}_{IJ}^{+})\varepsilon_{jk}Y^{ij\,I}-X_{-}^{I}\mathcal{F}_{IA}^{+}\varepsilon_{jk}B^{ij}_{+}\Big]\epsilon^k_{+}-2\mathrm{i}\eta_{+}^i\label{zeta_Q+},\\
\delta \zeta_{-}^{V i}=\,&e^{\mathcal{K}}\slashed{D}e^{-\mathcal{K}}\epsilon_{+}^{i}+
2\slashed{\mathcal{A}}\epsilon_{+}^{i}-\frac{1}{2}\mathrm{i}\mathcal{F}_{ab}^{+}\gamma^{ab}\epsilon_{-}^{i}\nonumber\\
\,&-\mathrm{i}e^{\mathcal{K}}\Big[(\mathcal{F}_{I}^{+}+X_{+}^{J}\mathcal{F}_{IJ}^{-})\varepsilon_{jk}Y^{ij\, I}-X_{+}^{I}\mathcal{F}_{IA}^{-}\varepsilon_{jk}B^{ij}_{-}\Big]\epsilon^k_{-}+2\mathrm{i}\eta_{-}^i\label{zeta_Q-},
\end{align}
where we have defined the composite quantities $\mathcal{F}_{ab}^{\pm}$ as,
\begin{align}
    \mathcal{F}_{ab}^{-}\,&\equiv e^{\mathcal{K}}\Big[\mathcal{F}_{I}^{-}{F}_{ab}^{I-}-X_{-}^{I}G_{ab I}^{-}\Big],\\
     \mathcal{F}_{ab}^{+}\,&\equiv e^{\mathcal{K}}\Big[\mathcal{F}_{I}^{+}{F}_{ab}^{I+}+X_{+}^{I}G_{ab I}^{+}\Big].
\end{align}
We see from equations  \eqref{zeta_Q+}--\eqref{zeta_Q-} that $\zeta_{\pm}^{Vi}$  transform as bare parameters under $S$-super-symmetry. This property makes them useful to construct $S$-invariant linear combinations.

To construct yet another $S$-supersymmetry compensator, we consider the following composite fermion using the fields of the hypermultiplets,
\begin{align}
\Bar{\zeta}_{i}^H\equiv\chi^{-1}\Omega_{\alpha\beta}A_{i}{}^{\alpha}\zeta^{\beta}.
\end{align}
The $Q$- and $S$-supersymmetry transformation of $\Bar{\zeta}_i^H$ is 
\begin{align}
\delta \Bar{\zeta}_{i}^{H}=\delta  \Bar{\zeta}_{i+}^{H}+\delta  \Bar{\zeta}_{i-}^{H},
\end{align}
where 
\begin{align}\label{zetah}
    \delta \Bar{\zeta}_{i+}^{H}=\,&-\mathrm{i}\chi^{-1}\Omega_{\alpha\beta}A_{i}{}^{\alpha}\slashed{D}A_{j}{}^{\beta}\epsilon_{-}^j+\varepsilon_{ij}\eta_{+}^j,\\[1mm]
    \delta \Bar{\zeta}_{i-}^{H}=\,&-\mathrm{i}\chi^{-1}\Omega_{\alpha\beta}A_{i}{}^{\alpha}\slashed{D}A_{j}{}^{\beta}\epsilon_{+}^j+\varepsilon_{ij}\eta_{-}^j .
\end{align}
We see from equations  \eqref{zetah} that $\Bar{\zeta}_{i\pm}^{H}$ also transform as bare parameters under $S$-super-symmetry.

We can write equations  \eqref{zetah} in a more useful form. As in \cite{LopesCardoso:2000qm}, we make the following decomposition,
\begin{align}
\chi^{-1}\Omega_{\alpha\beta}A_{i}{}^{\alpha}D_{\mu}A_{j}{}^{\beta}=\frac{1}{2}k_{\mu}\varepsilon_{ij}+k_{\mu ij},
\end{align}
where $k_{\mu}$ is defined as
\begin{align}
    k_\mu=\chi^{-1}(\partial_\mu -2b_\mu)\chi. 
\end{align} 
With this decomposition, equations~\eqref{zetah} become,
\begin{align}
      \delta \Bar{\zeta}_{i+}^{H}=\,&-\frac{\mathrm{i}}{2}\slashed{k}\varepsilon_{ij}\epsilon_{-}^j-\mathrm{i}\slashed{k}_{ ij}\epsilon_{-}^j+\varepsilon_{ij}\eta_{+}^j, \label{zetahplus}\\
    \delta \Bar{\zeta}_{i-}^{H}=\,&-\frac{\mathrm{i}}{2}\slashed{k}\varepsilon_{ij}\epsilon_{+}^j-\mathrm{i}\slashed{k}_{ ij}\epsilon_{+}^j+\varepsilon_{ij}\eta_{-}^j \label{zetahminus}.
\end{align}
We will also need variation of $D_{\mu}\Bar{\zeta}_{i\pm}^{H}$ in the following. We have,
\begin{align}\label{varDzeta}
    \delta (D_{\mu}\Bar{\zeta}_{i \pm}^{H})=\,& -\frac{\mathrm{i}}{2}\mathcal{D}_{\mu}\left(\chi^{-1}\mathcal{D}_{\nu}\chi\right)\,\,\varepsilon_{ij}\gamma^{\nu}\epsilon_{\mp}^j-\mathrm{i}\mathcal{D}_{\mu}k_{\nu ij}\,\gamma^{\nu}\,\epsilon_{\mp}^j\nonumber\\
    \,&-\frac{1}{32}\chi^{-1/2} \slashed{\mathcal{D}}(\chi^{1/2}T_{ab}^{\pm})\gamma^{ab}\gamma_{\mu}\varepsilon_{ij}\epsilon_{\pm}^j-\frac{1}{32}\slashed{k}_{ij}\, T_{ab}^{\pm}\gamma^{ab}\gamma_{\mu}\epsilon^j_{\pm}\nonumber\\
    \,& +\varepsilon_{ij}\left(-\mathrm{i} \,f_{\mu}{}^{a}\gamma_{a}\epsilon_{\mp}^{j}+\frac{\mathrm{i}}{8}R(\mathcal{V})_{ab}^{\mp j}{}_{k}\, \,\gamma^{ab}\gamma_{\mu} \epsilon^{k}_{\mp}\mp\frac{\mathrm{i}}{4}R(A)^{\mp}_{ab}\gamma^{ab}\gamma_{\mu}\epsilon_{\mp}^{j}\right)\nonumber\\
    \,&
    +\left(\frac{1}{4}\chi^{-1}\slashed{\mathcal{D}}\chi
\varepsilon_{ij}+\frac{1}{2}\slashed{k}_{ij}\right)\,\gamma_{\mu}\eta^{j}_{\pm},
\end{align}
where we have used \eqref{zetaalpha} and \eqref{transzeta}.
\subsection{Killing spinor equations}
In this subsection, we evaluate all the fermionic supersymmetry variations. By setting them to zero, we derive the resulting constraints for fully supersymmetric solutions. We start with the variation of the composite hypermultiplet fermion $\zeta_{i\pm}^{H}$. In contrast to the Lorentzian supergravity, the chiral and anti-chiral spinors are independent, so we consider their variations separately. For the chiral spinor, we use the following $S$-invariant linear combination and demand its $Q$-supersymmetry variation to vanish,
\begin{align}\label{zeta_plus_const}
    \delta \Bar{\zeta}_{i+}^{H}-\frac{\mathrm{i}}{2}\varepsilon_{ij}\delta\zeta_{+}^{V j}\overset{!}{=}0.
\end{align}
This gives us the following constraints,
\begin{align}
\mathcal{A}_{\mu}+\frac{1}{2}k_{\mu}\,&=\frac{1}{2}e^{\mathcal{K}}\mathcal{D}_{\mu}e^{-\mathcal{K}}\label{A+},\\
\mathcal{F}_{ab}^{-}\,&\equiv e^{\mathcal{K}}\left[\mathcal{F}_{I}^{-}{F}_{ab}^{-I}-X_{-}^{I}G_{ab I}^{-}\right]=0,\label{curlf-}\\
k_{\mu ij}\,&=0,\\
(\mathcal{F}_{I}^{-}+X_{-}^{J}\mathcal{F}_{IJ}^{+})Y^{ij \, I}\,&=X_{-}^{I}\mathcal{F}_{IA}^{+}B_{+}^{ij}\label{B+}.
\end{align}
 For the anti-chiral spinor, we take the linear combination,
\begin{align}\label{zeta_minus_const}
    \delta \Bar{\zeta}_{i-}^{H}+\frac{\mathrm{i}}{2}\varepsilon_{ij}\delta\zeta_{-}^{V j}\overset{!}{=}0,
\end{align}
which yields the following constraints,
\begin{align}
\mathcal{A}_{\mu}-\frac{1}{2}k_{\mu}\,&=-\frac{1}{2}e^{\mathcal{K}}\mathcal{D}_{\mu}e^{-\mathcal{K}},\label{A-}\\
\mathcal{F}_{ab}^{+}\,&\equiv e^{\mathcal{K}}\left[\mathcal{F}_{I}^{+}{F}_{ab}^{+I}+X_{+}^{I}G_{ab I}^{+}\right]=0,\label{curlf+}\\
k_{\mu ij}\,&=0,\\
(\mathcal{F}_{I}^{+}+X_{+}^{J}\mathcal{F}_{IJ}^{-})Y^{ ij\, I}\,&=X_{+}^{I}\mathcal{F}_{IA}^{-}B_{-}^{ij}\label{B-}.
\end{align}
Combining \eqref{A+} and \eqref{A-}, we get
\begin{align}\label{kappa chi}
    \mathcal{D}_{\mu}(e^{\mathcal{K}}\chi)=0=\mathcal{A}_{\mu}.
\end{align}

Next, we consider the variation of the vector multiplet gauginos $\Omega_{\pm}^{Ii}$. We consider $S$-invariant combinations and set their $Q$-supersymmetry variations to zero,
\begin{align}\label{Omega}
    \delta \Omega_{\pm}^{Ii}\mp \mathrm{i}X_{\pm}^I \delta\zeta_{\pm}^{Vi}\overset{!}{=}0.
\end{align}
This gives,
    \begin{align}
        {F}_{ab}^{I\pm}=\,&\frac{1}{4}X_{\pm}^{I}T_{ab}^{\pm},\label{fab}\\
        Y^{ij\,I}=\,&0,\label{Y}\\
       \mathcal{D}_{\mu}(e^{\mathcal{K}/2}X_{\pm}^{I})=\,&0\label{X}.
    \end{align}
Putting \eqref{Y} back in \eqref{B+} and \eqref{B-}, we see that the (anti-)chiral multiplet field $B^{ij}_{\pm}$ vanishes for fully supersymmetric solutions,
\begin{align}
    B^{ij}_{\pm}=0.
\end{align}

Combining \eqref{fab} with \eqref{curlf-} and \eqref{curlf+}, we obtain the dual electromagnetic field strength $G_{abI}$ in terms of $\mathcal{F}_{I}^{\pm}$ and $T_{ab}^{\pm}$, 
\begin{align}
    G^{\mp}_{abI}=\,&\pm\frac{1}{4}\mathcal{F}_{I}^{\mp}T_{ab}^{\mp}\label{gab1}.
\end{align}
We will make use of this when analyzing supersymmetric solutions carrying electromagnetic charges.

With the constraints obtained so far, the transformation of $\zeta_{\pm}^{Vi}$ simplifies to, 
\begin{align}
    \delta \zeta_{\pm}^{Vi}=\mp e^{\mathcal{K}}\slashed{\mathcal{D}}e^{-\mathcal{K}}\epsilon_{\mp}^{i}~\mp2\operatorname{i}\eta_{\pm}^{i}.
\end{align}
Next, we consider the variation of the hypermultiplet fermion $\zeta_{\pm}^{\alpha}$. We demand the variation of the following $S$-invariant linear combinations to vanish under $Q$-supersymmetry,
\begin{align}
    \delta \zeta^{\alpha}_{\pm}\mp\frac{\operatorname{i}}{2}A_{i}^{\alpha}\delta\zeta_{\pm}^{Vi}\overset{!}{=}0.
\end{align}
This, together with the condition \eqref{kappa chi}, yields
\begin{align}
    \mathcal{D}_{\mu}(\chi^{-1/2}A_{i}^{\alpha})=0.
\end{align}

Next, we consider the background (anti-) chiral multiplet spinor ${\Psi}^{i}_{\pm}$. We construct the following $S$-invariant combinations and set their $Q$-supersymmetry variations to zero,
\begin{align}
\delta {\Psi}^{i}_{\pm}\mp 2\mathrm{i}{A}_{\pm}\delta \zeta_{\pm}^{Vi}\overset{!}{=}0.
\end{align}
From this, we obtain the conditions
\begin{align}
   \mathcal{D}_{\mu}(e^{\mathcal{K}}{A}_{\pm})=\,&0,\label{A}\\
  {F}_{ab}^{\pm}=\,&0.\label{cfab}
\end{align}
The constraints on $X_{\pm}^{I}$ and $A_{\pm}$  in \eqref{X} and \eqref{A} respectively also constrain $\mathcal{F}^{\pm}_{I}$. We have, 
 \begin{align}
     \mathcal{D}_{\mu}(e^{\mathcal{K}/2}\mathcal{F}_{I}^{\pm})=\,&e^{\mathcal{K}/2}\mathcal{D}_{\mu}\mathcal{F}_{I}^{\pm}+\mathcal{F}_{I}^{\pm}\mathcal{D}_{\mu}e^{\mathcal{K}/2}.
 \end{align}
 Derivatives $\mathcal{D}_{\mu}\mathcal{F}_{I}^{\pm}$ can be expanded as
 \begin{align}
\mathcal{D}_{\mu}\mathcal{F}_{I}^{\pm}=\mathcal{F}_{IJ}^{\pm}\mathcal{D}_{\mu}X^{J}_{\pm}+\mathcal{F}_{IA}^{\pm}\mathcal{D}_{\mu}A_{\pm}.
 \end{align}
Using the homogeneity property of $\mathcal{F}^{\pm}_{I}$ \eqref{homo} we have 
 \begin{align}
     \mathcal{D}_{\mu}(e^{\mathcal{K}/2}\mathcal{F}_{I}^{\pm})=\mathcal{F}_{IJ}^{\pm}\mathcal{D}_{\mu}(e^{\mathcal{K}/2}X_{\pm}^{J})+\mathcal{F}_{IA}^{\pm}e^{-\mathcal{K}/2}\mathcal{D}_{\mu}(e^{\mathcal{K}}A_{\pm})=0\label{cnstrprepot}.
 \end{align}

Next, we turn to the variations of the Weyl multiplet fermions. Due to the fact that the transformations of the gauge fields cannot be consistently set to zero, as they can always be shifted using gauge symmetries, we need to consider only covariant spinor fields. We start with the variation of the covariant auxiliary field $\chi^{i}_{\pm}$. From \eqref{weylvari}, we have,
\begin{align}
    \delta \chi_{\pm}^i= -\frac{\mathrm{i}}{6}\gamma_b \mathcal{D}_{a}T^{ab\mp}\epsilon_{\mp}^{i}+\frac{1}{6}R(\mathcal{V})_{ab}{}{}^{i\mp}{}_{j}\gamma^{ab}\epsilon_{\pm}^{j}\mp\frac{1}{3}R(A)_{ab}^{\mp}\gamma^{ab}\epsilon_{\pm}^{i}+D\epsilon_{\pm}^{i}+\frac{1}{24}T_{ab}^{\mp}\gamma^{ab}\eta_{\pm}^{i}.
\end{align}
An $S$-invariant combination can be readily constructed using  $\chi_{\pm}^i$ and $\zeta_{\pm}^{Vi}$, whose $Q$-variation we need to set to zero,
\begin{align}
    \delta\chi_{\pm}^{i}\mp\frac{\mathrm{i}}{48}T_{ab}^{\mp}\gamma^{ab}\delta \zeta_{\pm}^{Vi}\overset{!}{=}0.
\end{align}
This yields the following conditions,
    \begin{align}
        D=\,&0\label{D=0},\\
        R(\mathcal{V})_{ab}{}{}^{i\pm}{}_{j}=\,&0,\label{zeroRV}\\
        R(A)^{\pm}_{ab}=\,&0,\label{RA=0}\\
        \mathcal{D}_{a} \left(e^{-\mathcal{K}/2}T^{ab\mp}\right)=\,&0.\label{Dtab=0}
    \end{align}
For constraints \eqref{zeroRV} and \eqref{RA=0}, we can take the solutions of the gauge fields as
\begin{align}
    \mathcal V_{\mu}{}^{i}{}_{j}=0=A_{\mu}.
\end{align}
Now using \eqref{fab}, \eqref{X}, \eqref{gab1}, \eqref{cnstrprepot}, and \eqref{Dtab=0}, we find that  field strengths are covariantly constant,
\begin{align}
    \mathcal{D}_{a}F^{I ab\pm}=0,~\qquad ~\mathcal{D}_{a}G_{I}^{ab\pm}=0.
\end{align}
The above conditions on the field strengths imply equations of motion \eqref{eom} and the Bianchi identities \eqref{bian}.

Next, we take the curvature $R(Q)_{ab}^{i}$ as the covariant spinor field and construct a suitable $S$-invariant linear combination. We demand its $Q$-variation to go to zero,
\begin{align}
 \delta R(Q)_{ab\pm}^{i}\mp \frac{\mathrm{i}}{32}T_{cd}^{\mp}\gamma^{cd}\gamma_{ab}\delta \zeta_{\pm}^{Vi}\overset{!}{=}0.
\end{align}
We get the following constraints,
\begin{align}
    \mathcal{R}(M)_{ab}{}^{cd}=\,&0,\label{RM}\\
    \mathcal{D}_{c
     }T_{ab}^{\pm}=\,&\frac{1}{2}e^{\mathcal{K}}\mathcal{D}_{e}e^{-\mathcal{K}}\Big(\delta^{e}_{c}T_{ab}^{\pm}+2\delta_{c[a}T_{b]}{}^{e\pm}-2\delta^{e}_{[a}T_{b]c}^{\pm}\Big)\label{dtab}.
\end{align}

Next, we turn to the variation of the background spinor ${\Lambda}^i$. Using the previous constraints, we see that it is $S$-invariant. Thus, we only need to set its $Q$-variation to zero,
\begin{align}
    \delta {\Lambda}^{i}_{\pm}={C}_{\pm}\epsilon^{i}_{\pm}+\frac{\mathrm{i}}{8}\left(\slashed{D}{A}_{\pm}T_{ab}^{\pm}+2 {A}_{\pm}\slashed{D}T_{ab}^{\pm}\right)\gamma^{ab}\epsilon^{i}_{\mp}\overset{!}{=}0.
\end{align}
The second term vanishes by virtue of \eqref{A}
and \eqref{Dtab=0}.  The non-trivial condition we get is
 \begin{align}\label{cplusminus=0}
     C_{\pm}=0.
 \end{align}
 The bosonic part of the expression of $C_{\pm}$ from
\eqref{eq:weyl-squared-components}
 in terms of the Weyl multiplet fields is, 
 \begin{align}
      {C}_\pm =&\,
  -64\,\mathcal{R}(M)^{cd\mp}{\!}_{ab}\,\mathcal{R}(M)_{cd}{\!}^{\mp \,ab}
  - 32\,R(\mathcal{V})^{\mp\,ab\,i}{}_j
  \,R(\mathcal{V})_{ab}{\!}^{\mp j}{}_i 
+16\, T^{\mp\,ab}D_a D^c T_{cb}{\!}^\pm.\label{Cab}
 \end{align}
Using \eqref{zeroRV} and \eqref{RM}, we see that the condition  \eqref{cplusminus=0} gives a condition on the derivative of $T_{ab}^{\pm}$, 
\begin{align}\label{ddtpm=0}
     T^{\mp\,ab}D_a D^c T_{cb}{\!}^\pm=0.
\end{align}
We can rewrite this as, 
\begin{align}\label{T split}
     T^{\mp\,ab}D_a D^c T_{cb}{\!}^\pm =T^{\mp\,ab}\mathcal{D}_a \mathcal{D}^c T_{cb}{\!}^\pm-f_{a}{}^{c}T^{ab\mp}T_{cb}{}^{\pm}=0.
\end{align}
The first term can be computed using \eqref{dtab}
which allows us to solve for $f_{\mu}{}^{a}$. We find, 
\begin{align}\label{Kgaugefield}
    f_{a}{}^{c}=-\frac{1}{2}\mathcal{D}_{a}(e^{\mathcal{K}}\mathcal{D}^c e^{-\mathcal{K}})+\frac{1}{4}(e^{\mathcal{K}}\mathcal{D}_{a}e^{-\mathcal{K}})(e^{\mathcal{K}}\mathcal{D}^{c}e^{-\mathcal{K}})-\frac{1}{8}\delta^{c}_{a}(e^{\mathcal{K}}\mathcal{D}^{e}e^{-\mathcal{K}})^2.
\end{align}

Two comments are in order here. 

First, in our analysis so far we have chosen the composite fermion $\zeta_{\pm}^{Vi}$ as the compensator for the $S$-transformation. The composite fermion
$\Bar{\zeta}_{i\pm}^{H}$ could also be used for this purpose. The constraints obtained using $\Bar{\zeta}_{i\pm}^{H}$ as compensator are not different from what we obtained above. In other words, the variations of the $S$-invariant combinations of the fermions with $\Bar{\zeta}_{i\pm}^{H}$ vanish once the above constraints are imposed. 

Second, we make a technical observation regarding covariant derivatives of  fermion fields \cite{LopesCardoso:2000qm}.  We note that a derivative of an arbitrary fermion can be written as the sum of the following terms: a bosonic expression times $\Bar{\zeta}_{i\pm}^{H}$, and a term proportional to the variation of the following $S$-invariant expression,
\begin{align}\label{derizeta}
    D_{\mu}\Bar{\zeta}_{i\pm}^{H}+\left(-\frac{1}{4}\chi^{-1}\slashed{{D}}\chi
\delta_i^j+\frac{1}{2}\slashed{k}_{ik}\varepsilon^{kj}\right)\,\gamma_{\mu}\Bar{\zeta}_{j\pm}^{H} .
\end{align}
If we demand the $Q$-variation of the above expression to vanish, then the variation of the covariant derivative of any fermion vanishes. Using \eqref{varDzeta}, one can show that the variation of \eqref{derizeta} vanishes upon using \eqref{Kgaugefield}.\footnote{Alternatively, equation \eqref{Kgaugefield} can be obtained by demanding the variation of \eqref{derizeta} vanishes.} Consequently, we do not need to consider variations of covariant derivatives of fermions separately.

So far, all constraints we have obtained are consistent with superconformal symmetries. To get to the Euclidean theory of interest, we need to impose gauge-fixing conditions. Our dilatation gauge fixing condition is,
\begin{align}
    \chi=\text{constant}.
\end{align}
The $D$-equation \eqref{Deom} evaluated on fully supersymmetric configurations takes the form,
\begin{align}
    e^{-\mathcal{K}}+\frac{1}{2}\chi=0 \implies  e^{-\mathcal{K}} = \text{constant}.
\end{align}
This in turn implies that the right hand side of \eqref{Kgaugefield} vanishes, i.e.,  
\be
f_{\mu}{}^{a}=0. 
\ee
Furthermore, constraint \eqref{dtab} now reduces to 
 \begin{align}\label{ConstantT}
     \mathcal{D}_{c}T_{ab}^{\pm}=0.
 \end{align}

 Our gauge fixing condition for special conformal transformation is,  
 \begin{align}
     b_{\mu}=0.
 \end{align}
 From \eqref{X}, we see that this choice implies that the vector multiplet scalars are constants,
 \begin{align}
     \partial_{\mu}X_{\pm}^{I}=0.
 \end{align}
In the gauge $b_{\mu}=0$, $R(\omega,e)_{\mu}{}^{a}$ and $R(\omega,e)$ can be interpreted as the standard Ricci tensor and Ricci scalar, respectively, which appear in the expression of $f_{\mu}{}^{a}$ given in \eqref{fmua}. Plugging  constraints \eqref{D=0} and \eqref{RA=0} in \eqref{fmua}, we obtain the components of the Ricci tensor in terms of the $T$-fields,
\begin{align}\label{RTT}
    R_{\mu\nu}=\frac{1}{16}T_{\mu\rho}^{-}T^{\rho}{}_{\nu}^{+}.
\end{align}
In particular, the Ricci scalar vanishes since it is a contraction of self-dual and anti-self-dual tensors. Furthermore,  the modified curvature \eqref{curlyR} reduces to the standard Weyl tensor, which vanishes by virtue of the constraint \eqref{RM}. Thus, we conclude that for fully supersymmetric configurations Weyl tensor  vanishes,
\begin{align}
C_{\mu\nu\rho\sigma}=0.
\end{align}
 
\subsection{\texorpdfstring{Euclidean $AdS_2\times S^2$ and Wald entropy}{Euclidean AdS2 x S2 and Wald entropy}}
In this subsection, we discuss a class of explicit solutions that satisfy the constraints obtained in the previous subsection. 
Since the Weyl tensor vanishes, the metric is conformally flat. Following \cite{LopesCardoso:2000qm}, we write the metric ansatz in the following form,
\begin{align}
g_{\mu\nu}=e^{2f+\mathcal{K}}\delta_{\mu\nu},
\end{align}
where $e^{2f}$ is the conformal factor and $e^{\mathcal{K}}$ is included to make the function $f$ scale independent. Our goal is to find the function $f$ by solving \eqref{ConstantT} and \eqref{RTT}.
The vanishing of the Ricci scalar implies 
\begin{align}
\delta^{\mu\nu}\partial_{\mu}\partial_{\nu}e^f=0,
\end{align}
i.e., $e^{f}$ is harmonic in four dimensions. We next compute the Ricci tensor components and compare with \eqref{RTT}. We get
\begin{align}
R_{\mu\nu}=\,&2\partial_{\mu}\partial_{\nu}f-2\partial_{\mu}f\partial_{\nu}f+\delta_{\mu\nu}(\partial_{\rho}f)^2=\frac{1}{16}T_{\mu\rho}^{+}\check{T}^{\rho}{}_{\nu}^{-}e^{-2f-\mathcal{K}}\label{ricci}.
\end{align}
where we have defined
\begin{align}
\check{T}^{\mu\nu}\equiv\delta^{\mu\rho}\delta^{\nu\sigma}T_{\rho\sigma}.
\end{align}

We now expand the covariant derivative in \eqref{ConstantT} to get (again, indices are raised by the flat metric),
\begin{align}
\partial_{\mu}T_{\nu\rho}^{\pm}=\,&2\partial_{\mu}f ~T_{\nu\rho}^{\pm}-2\partial_{[\nu}f~T_{\rho]\mu}^{\pm}+2\delta_{\mu[\nu}T_{\rho]\sigma}^{\pm}\partial^{\sigma}f\label{DT}.
\end{align}
From \eqref{DT} and the (anti-)self-duality property of $T^{\pm}_{\mu \nu}$ it follows that, 
\begin{align}\label{harmonic}
    \partial_{\mu} \check{T}^{\mu\nu\pm}=0.
\end{align}

We now focus on ``time-independent'' solutions, which correspond to field configurations where all fields are independent of the Euclidean time $\tau$.
In that case, \eqref{harmonic} splits into,
\begin{align}
\partial_{\hat{a}}\check{T}^{\hat{a}\hat{b}\pm}=0, ~~\partial_{\hat{a}}\check{T}^{\tau\hat{a}\pm}=0,
\end{align}
where $\hat{a}, \hat{b}, \hat{c}$ denotes world spatial indices and $\tau$ is the Euclidean time. 
These equations allow us to write $T_{\mu\nu}^{\pm}$ as,
\begin{align}
{T}_{\hat{a}\hat{b}}^{+}&=\varepsilon_{\hat{a}\hat{b}\hat{c}}\partial^{\hat{c}}\varphi,& {T}_{\tau \hat{a}}^{+}& = \partial_{\hat{a}}\varphi\label{T+},\\
{T}_{\hat{a}\hat{b}}^{-}&=\varepsilon_{\hat{a}\hat{b}\hat{c}}\partial^{\hat{c}}\xi, & {T}_{\tau \hat{a}}^{-}&= -\partial_{\hat{a}}\xi\label{T-}.
\end{align}
Here $\varphi$ and $\xi$ are real harmonic functions and $\varepsilon_{\hat{a}\hat{b}\hat{c}}$ is the Levi-Civita tensor of the three-dimensional flat space. We note that the constraint $T_{\mu\nu}^{+}\check{T}^{\mu\nu-}=0$ is satisfied.

We also split the Ricci tensor \eqref{ricci} into its spatial and temporal components as,
\begin{align}
R_{\hat{a}\hat{b}}=\,&2\partial_{\hat{a}}\partial_{\hat{b}}f-2\partial_{\hat{a}}f\partial_{\hat{b}}f+\delta_{\hat{a}\hat{b}}(\partial_{\hat{c}}f)^2=\frac{1}{16}\Big(T^{+}_{\hat{a}\hat{c}}\check{T}^{\hat{c}}{}_{\hat{b}}^{-}+T^{+}_{\hat{a}\tau}\check{T}^{\tau}{}_{\hat{b}}^{-}\Big)e^{-2f-\mathcal{K}}\label{Rab},\\
R_{\tau\tau}=\,&\partial_{\hat{a}}f\partial^{\hat{a}}f=\frac{1}{16}T_{\tau\hat{a}}^{+}\check{T}^{\hat{a}}{}_{\tau}^{-}e^{-2f-\mathcal{K}}\label{Rtt}.
\end{align}
We now plug \eqref{T+} and \eqref{T-} in \eqref{Rtt}, to get
\begin{align}\label{phi-xi-f}
\partial_{\hat{a}}\varphi\partial^{\hat{a}}\xi=16 e^{2f+\mathcal{K}}\partial_{\hat{a}}f\partial^{\hat{a}}f.
\end{align}

To solve the above differential equation, we consider an ansatz (a family of solutions) parametrized by constant real parameters  $z_1$ and $z_2$,
\begin{align}
    \varphi=\,&4 z_1e^{f+\mathcal{K}/2},\\
    \xi=\,&4z_2
    e^{f+\mathcal{K}/2}.
\end{align}
Putting the above ansatz back into \eqref{Rtt}, we get
\begin{align}
   (z_1z_2-1) \partial_{\hat{a}}f\partial^{\hat{a}}f=0
\end{align}
For non-flat solutions, i.e., $\partial_{\hat{a}}f\neq0$, we must have
\begin{align}
    z_1z_2=1.
\end{align}
Inserting the above ansatz into \eqref{Rab}, we get a differential equation for the function $f$,
\begin{align}\label{finalf}
e^f\partial_{\hat{a}}\partial_{\hat{b}}e^f=3\partial_{\hat{a}}e^f\partial_{\hat{b}}e^f-\delta_{\hat{a}\hat{b}}(\partial_{\hat{c}}e^f)^2.
\end{align}

We solve this differential equation for spherically symmetric solutions, $f=f(\rho)$,  where  $\rho$ is the radial coordinate $\rho^2=x^{\hat{a}}x_{\hat{a}}$ . Equation \eqref{finalf} now takes the form,
\begin{align}
    \frac{df}{d\rho}\left(\frac{df}{d\rho} + \frac{1}{\rho}\right)=0.
\end{align}
This equation is solved as
\begin{align}
    \frac{df}{d\rho}=-\frac{1}{\rho}\implies e^f=\frac{c}{\rho},
\end{align}
where $c$ is a real constant. As a result, the metric reads
\begin{align}
    ds^2=\frac{c^2 e^{\cK}}{\rho^2}(d\tau^2+d\rho^2+\rho^2(d\theta^2+\sin^2{\theta}d\phi^2)).
\end{align}
By redefining the radial variable as $\rho=\frac{c^2}{r}$, we get a standard form of Euclidean $AdS_2\times S^2$,
\begin{align}\label{Ads}
ds^2=e^{\cK}\Big[\frac{r^2}{c^2}d\tau^2+\frac{c^2}{r^2}dr^2+c^2(d\theta^2+\sin^2{\theta}d\phi^2)\Big].
\end{align}
This metric describes the near-horizon geometry of the Euclidean extremal  Reissner-Nordstr\"om black hole.
The magnetic and electric charges carried by the solution can be defined as
\begin{align}
    p^{I}=\frac{1}{4\pi}\,&\oint_{S^2}F^{I}=\frac{1}{4\pi}\int d\theta d\phi~ F_{\theta\phi}^{I},\\
    q_{I}=\frac{1}{4\pi}\,&\oint_{S^2}G_{I}=\frac{1}{4\pi}\int d\theta d\phi~ G_{\theta\phi I}.
\end{align}

Using the constraints we obtained above we  compute
\begin{align}
&F_{\theta\phi}^{I}= F_{\theta\phi}^{+I}+F_{\theta\phi}^{-I}=\frac{1}{4}\Big(X_{+}^{I}T_{\theta\phi}^{+}+X_{-}^{I}T_{\theta\phi}^{-}\Big),\\
 & G_{\theta\phi I}= G_{\theta\phi I}^{+}+G_{\theta\phi I}^{-}=\frac{1}{4}\Big(\mathcal{F}_{I}^{-}T_{\theta\phi}^{-}-\mathcal{F}_{I}^{+}T_{\theta\phi}^{+}\Big).
\end{align}
Using \eqref{T+} and \eqref{T-}, the auxiliary $T$-field which take the form  
\begin{align}
    T_{\theta\phi}^{+}=-4z_1\,c \, e^{\mathcal K/2}\sin{\theta},\, \quad \, T_{\theta\phi}^{-}=-4z_2\,c \, e^{\mathcal K/2}\sin{\theta}
\end{align}
Integrating the field strengths over the $S^2$, 
we obtain the magnetic and
electric charges
(with orientation fixed by $\varepsilon_{\hat 1 \hat 2 \hat 3 }=1$ and $\hat x^a = \rho( \sin \theta \cos \phi, \sin \theta \sin \phi, \cos \theta))$,
\begin{align}
    p^I=\,&-ce^{\mathcal{K}/2}(z_1X_{+}^{I}+z_2X_{-}^{I})\label{mag},\\[1mm]
    q_{I}=\,&-ce^{\mathcal{K}/2}(z_2\mathcal{F}^{-}_{I}-z_1\mathcal{F}^{+}_{I}).\label{elec}\end{align}
The minus signs can be absorbed by redefining the parameters as $\tilde z_{1}=-z_1$ and $\tilde z_2=-z_2$. Inverting \eqref{mag} and \eqref{elec}, in terms of the new parameters, we find
\begin{align}
 Z_{+}\equiv \,&e^{\mathcal{K}/2}(X_{+}^{I}q_{I}+\mathcal{F}_{I}^{+}p^{I})=c\tilde z_2, \\[1mm]
    Z_{-}\equiv \,&e^{\mathcal{K}/2}(\mathcal{F}_{
    I}^{-}p^{I}-X_{-}^{I}q_{I})=c\tilde z_1.
\end{align}
Now we can write the Euclidean version of the attractor equations,
\begin{align}
   e^{-\mathcal{K}/2} p^I=\,&(Z_{-}X_{+}^{I}+Z_{+}X_{-}^{I}),\label{Xattractor}\\[1mm]
       e^{-\mathcal{K}/2} q_{I}=\,&(Z_{+}\mathcal{F}^{-}_{I}-Z_{-}\mathcal{F}^{+}_{I}).\label{Fattractor}
\end{align}
We note that this result is covariant with respect to electric-magnetic duality, wherein 
\be
 \begin{pmatrix} 
     X_\pm^I \\[1mm]
      \mp\mathcal{F}^\pm_I
    \end{pmatrix}
    \ee
form symplectic pairs~\cite{deWit:2017cle}.
 
 The chiral multiplet fields ${A}_{\pm}$ become,
 \begin{align}
  {A}_{+}=\,& 4~T_{\theta\phi}^{-}T^{\theta\phi-}=64~Z^{-2}_{-}e^{-\mathcal{K}},\\[1mm]
{A}_{-}=\,& 4~T_{\theta\phi}^{+}T^{\theta\phi+}=64 ~Z^{-2}_{+}e^{-\mathcal{K}}. 
 \end{align}
Introducing the rescaled variables, 
 \begin{align}
     &Y_{+}^{I}=e^{\cK/2}Z_{-}X_{+}^{I}, &  & Y_{-}^{I}=e^{\cK/2}Z_{+}X_{-}^{I},\\[1mm]
     &\Upsilon_{+}=e^{\cK}Z_{-}^2 A_{+} = 64,&   & \Upsilon_{-}=e^{\cK}Z_{+}^2 A_{-}= 64,
 \end{align}
and using the homogeneity property of the prepotential functions $\mathcal{F}^{\pm}$, we can write the attractor equations in terms of the rescaled variables as,
     \begin{align}
    p^I=\,&(Y_{+}^{I}+Y_{-}^{I}),\\[1mm]
        q_{I}=\,&(\mathcal{F}^{-}_{I}(Y_{-}, 64)-\mathcal{F}^{+}_{I}(Y_{+}, 64)).\,
\end{align}
This form resembles the Lorentzian attractor equations in \cite{LopesCardoso:2000qm, Mohaupt:2000mj}, where the moduli fields are completely determined by the electric and magnetic charges at the horizon. For the prepotential functions
$\mathcal{F}^{+}$ and $\mathcal{F}^{-}$, these equations determine $Y_{\pm}^{I}$ in terms of the electric and magnetic charges.

\subsection*{Wald entropy}

We can now compute the Wald entropy of the single center black hole whose near horizon geometry correspond to the above discussed $AdS_2 \times S^2$.  Wald entropy for a black hole is given as \cite{Mohaupt:2000mj},
\begin{align}\label{wald}
    \mathcal{S}=-2\pi\oint d^2x \sqrt{h}~ \varepsilon_{ab}\varepsilon_{cd}\frac{\partial \mathcal{L}}{\partial R_{abcd}},
\end{align}
where $h$ is the determinant of the induced metric on the horizon and $\varepsilon_{ab}$ is the binormal at the horizon.
The relevant part of the Lagrangian \eqref{action} for the computation of Wald entropy reads,
\begin{align}
    8\pi \mathcal{L}=\,&\chi\left(\frac{1}{6} R+\frac{1}{2}D\right)-\left(\mathcal{F}_{I}^{+}X_{-}^{I}+\mathcal{F}_{I}^{-}X_{+}^{I}\right)\left(\frac{1}{6}R-D\right)\nonumber\\
   \,&  -\frac{1}{2}\left(\mathcal{F}_{A}^{+}C_{+}+\mathcal{F}_{A}^{-}C_{-}\right)+\frac{1}{4}\left(\mathcal{F}_{AA}^{-}F_{ab}^{+}F^{+ab}+\mathcal{F}_{AA}^{+}F_{ab}^{-}F^{-ab}\right)\nonumber\\
   \,&
   +\frac{1}{2}\Bigg[\mathcal{F}_{AI}^{+}F^{ab-}\Big(F_{ab}^{-I}-\frac{1}{4}X_{-}^{I}T_{ab}^{-}\Big)+\mathcal{F}_{AI}^{-}F^{ab+}\Big(F_{ab}^{+I}-\frac{1}{4}X_{+}^{I}T_{ab}^{+}\Big)\Bigg]+\cdots. 
\end{align}
The ellipsis contains terms that do not depend on the Riemann tensor.  The $8\pi$ factor on the left-hand side serves as a normalization factor to obtain the correct factor $(16\pi G_{N})^{-1}$ in the Einstein–Hilbert term after choosing $e^{\mathcal{K}} = G_{N}$, where $G_{N}$ is Newton’s gravitational constant.

For the metric \eqref{Ads}, non-vanishing components of the binormal are,
\be      
\varepsilon_{\tau r}= - \varepsilon_{r \tau} = e^{\cK},
\ee 
which implies,     
\be       \varepsilon_{\mu\nu}\varepsilon^{\mu\nu}=2\varepsilon_{\tau r}\varepsilon^{\tau r}=2(\varepsilon_{\tau r})^2g^{\tau \tau }g^{rr}=2.
\ee
To compute Wald entropy, we need to evaluate the 
derivative $
 8\pi\frac{\partial \mathcal{L}}{\partial R_{abcd}}$ in the fully supersymmetric $AdS_2 \times S^2$ background. We find,
\begin{align}
    8\pi\frac{\partial \mathcal{L}}{\partial R_{abcd}}=\,&-\frac{1}{2} e^{-\mathcal{K}} \frac{\partial R}{\partial R_{abcd}}-\frac{1}{2}\left(\mathcal{F}_{A}^{+}\frac{\partial C_{+}}{\partial R_{abcd}}+\mathcal{F}_{A}^{-}\frac{\partial C_{-}}{\partial R_{abcd}}\right)\nonumber\\
    \,&+\frac{1}{2}\left(\mathcal{F}_{AA}^{-}\frac{\partial F_{ef}^{+}}{\partial R_{abcd}}F^{+ef}+\mathcal{F}_{AA}^{+}\frac{\partial F_{ef}^{-}}{\partial R_{abcd}}F^{-ef}\right)\nonumber\\
    \,&+\frac{1}{2}\left[\mathcal{F}_{AI}^{+}\frac{\partial F_{ef}^{-}}{\partial R_{abcd}}\left(F_{ef}^{-I}-\frac{1}{4}X_{-}^{I}T_{ef}^{-}\right)+\mathcal{F}_{AI}^{-}\frac{\partial F_{ef}^{+}}{\partial R_{abcd}}\left(F_{ef}^{+I}-\frac{1}{4}X_{+}^{I}T_{ef}^{+}\right)\right]. \label{derivative}
\end{align}
In the fully supersymmetric background  $F_{ab}^{\pm}=0$ and $F_{ab}^{\pm I}=\frac{1}{4}X_{\pm}^{I}T_{ab}^{\pm}$. 
Expression \eqref{derivative} simplifies to,
\begin{align}
   \frac{\partial \mathcal{L}}{\partial R_{abcd}}=\,&-\frac{1}{16\pi} e^{-\mathcal{K}} \frac{\partial R}{\partial R_{abcd}}-\frac{1}{16\pi}\left(\mathcal{F}_{A}^{+}\frac{\partial C_{+}}{\partial R_{abcd}}+\mathcal{F}_{A}^{-}\frac{\partial C_{-}}{\partial R_{abcd}}\right).
\end{align}
In this expression, the first term upon contracting with $\varepsilon_{ab}\varepsilon_{cd}$ becomes,
\begin{align}\label{Rderivative}
-\frac{1} {16\pi}e^{-\mathcal{K}} \varepsilon_{ab}\varepsilon_{cd}g^{ac}g^{db}=-\frac{1}{8\pi} e^{-\mathcal{K}}. 
\end{align}
For the second term, using the expression of $C_{\pm}$ given in \eqref{eq:weyl-squared-components} we get, 
\begin{align}\label{Cderivative}
\varepsilon_{ab}\varepsilon_{cd}\frac{\partial C_{\pm}}{\partial R_{abcd}}=-8\varepsilon_{ab}\varepsilon_{cd}\delta^{ac}T^{\mp be}T^{\pm d}{}_{e}=-16 T^{\mp\tau r}T_{ \tau r}^{\pm}=256 c^{-2} e^{-\mathcal{K}}.
\end{align}
Putting \eqref{Rderivative} and \eqref{Cderivative} in \eqref{wald}, we get Wald entropy,
\begin{align}\label{euclideanentropy}
    \mathcal{S}=\pi\left[ Z_{+}Z_{-}+128(\mathcal{F}_{A}^{+}+\mathcal{F}_{A}^{-})\right], \qquad \qquad {A}_{\pm}=64 ~Z^{-2}_{\mp}e^{-\mathcal{K}},
\end{align}
where the first term $Z_{+}Z_{-}=c^2$ corresponds to the Bekenstein-Hawking entropy of the black hole, and the second term corresponds to the $R^2$ correction to the black hole entropy.
As in the Lorentzian case, the entropy is entirely determined by the charges, once the attractor equations \eqref{Xattractor} and \eqref{Fattractor} are imposed.

\subsection{Flat space}

Now we discuss other solutions to \eqref{phi-xi-f} with the following configurations,
\begin{align}
\varphi&=0, \qquad \xi \neq 0, \label{case1}\\
\text{or}\qquad \, \varphi&\neq 0, \qquad \xi=0. \label{case2}
\end{align}
These solutions require,
\begin{align}
\partial_{\hat{a}}f=0,
\end{align}
which renders the metric flat. However, the remaining fields are non-trivial. It is important to note that in the Euclidean signature $T_{ab}^{+}$ and $T_{ab}^{-}$ are independent real fields. There can be configurations where only one of the  $T_{ab}^{+}$ or $T_{ab}^{-}$ vanishes. The two solutions \eqref{case1}--\eqref{case2} correspond to such configurations. This is in sharp contrast to the Lorentzian case, where $T_{ab}^{+}$ and $T_{ab}^{-}$ 
are related by complex conjugation, and both vanish simultaneously.

Let us consider the first case, i.e.,  $\varphi = 0$. We have,
\begin{align}
    T_{\hat{a}\hat{b}}^{+}=0, ~~T_{\tau \hat{a}}^{+}= 0,\\
    T_{\hat{a}\hat{b}}^{-}=\varepsilon_{\hat{a}\hat{b}\hat{c}}\partial^{\hat{c}}\xi, ~~T_{\tau \hat{a}}^{-}= -\partial_{\hat{a}}\xi.
\end{align}
together with the field strengths,
        \begin{align}
        {F}_{ab}^{I+}&=0, &
        G^{+}_{abI}&=0,& 
        A_{-}&=0,&
        {F}_{ab}^{I-}&=\frac{1}{4}X_{-}^{I}T_{ab}^{-},&
        G^{-}_{abI}&=\frac{1}{4}\mathcal{F} _{I}^{-}T_{ab}^{-}.
    \end{align}
The fields $X_{\pm}^{I}$ take arbitrary constant values, and $\xi$ is at most linear in $\hat x^
a$, so that $T^-_{ab}$ is constant.

For the second case $\xi = 0$. We have 
\begin{align}
      T_{\hat{a}\hat{b}}^{-}=0, ~~T_{\tau \hat{a}}^{-}= 0,\\
T_{\hat{a}\hat{b}}^{+}=\varepsilon_{\hat{a}\hat{b}\hat{c}}\partial^{\hat{c}}\varphi, ~~T_{\tau \hat{a}}^{+}= \partial_{\hat{a}}\varphi,
\end{align}
together with the field strengths,
        \begin{align}
        {F}_{ab}^{I-}&=0, &
        G^{-}_{abI}&=0,&
        A_{+}&=0,&
        {F}_{ab}^{I+}&=\frac{1}{4}X_{+}^{I}T_{ab}^{+},&
        G^{+}_{abI}&=-\frac{1}{4}\mathcal{F} _{I}^{+}T_{ab}^{+}.
    \end{align}
The fields $X_{\pm}^{I}$ take arbitrary constant values, and $\varphi$ is at most linear in $\hat x^
a$, so that $T^+_{ab}$ is constant.

These solutions are crucial to understanding the new attractor mechanism in Euclidean supergravity. We will explore the link in our future work. 

\section{\texorpdfstring{Half BPS solutions}{Half BPS solutions}}\label{sec:half}
 In this section, we require that the supersymmetry transformation of the fermions vanish, but now only for half the total number of supersymmetries. We thus put a constraint on the supersymmetry parameters, 
 \begin{align}\label{condembedd}
    \gamma^0\epsilon_{+}^{i}=\mathrm{i}h \epsilon_{-}^{i}, 
\end{align}
where $h$ is real. The constraint makes half of the total number of supersymmetry parameters depend on the other half. We call this the embedding condition. The embedding condition
is obviously inconsistent with local four-dimensional rotational invariance as it singles out $\gamma^0$. The embedding condition presupposes that we would eventually impose a gauge condition that breaks the four-dimensional local rotational invariance. We will indeed do so later in our analysis. 

With this embedding condition, we perform the fermionic variations along the lines of the full supersymmetry analysis. Throughout the analysis, we take the same $S$-invariant linear combinations as chosen in the previous section and set their $Q$-variations to zero. 

We start with the hypermultiplet fermions $\Bar{\zeta}_{i\pm}^{H}$. Setting
\begin{align}
    \delta \Bar{\zeta}_{i\pm}^{H}\mp\frac{\mathrm{i}}{2}\varepsilon_{ij}\delta\zeta_{\pm}^{V j}\overset{!}{=}0,
\end{align}
gives (in the following $a = (0, p)$ with $p=1,2,3$ refer to the tangent space),
\begin{align}
    \mathcal{A}_{0}=\,&0,\\
    \mathcal{A}_{p}=\,&\frac{1}{2}\left(h\mathcal{F}_{p0}^{-}+h^{-1}\mathcal{F}_{p0}^{+}\right),\\
    k_{\mu ij}=\,&0,\\
    \mathcal{D}_{0}(\chi e^{\cK})=\,&0,\\
    \mathcal{D}_{p}(\chi e^{\mathcal{K}})=\,& \chi e^{\mathcal{K}}\left(h\mathcal{F}^{-}_{p0}-h^{-1}\mathcal{F}_{p0}^{+}\right),\\
    \left(\mathcal{F}_{I}^{-}+X_{-}^{J}\mathcal{F}_{IJ}^{+}\right)Y^{ij \, I}=\,&X_{-}^{I}\mathcal{F}_{IA}^{+}B_{+}^{ij},\\
\left(\mathcal{F}_{I}^{+}+X_{+}^{J}\mathcal{F}_{IJ}^{-}\right)Y^{ ij \, I}=\,&X_{+}^{I}\mathcal{F}_{IA}^{-}B_{-}^{ij}.
\end{align}
Using these equations, the transformations of $\zeta_{i\pm}^{V}$ simplify to 
    \begin{align}
\delta\zeta_{+}^{V i}=\,&-\chi^{-1}\slashed{D}\chi \epsilon_{-}^{i}-2\mathrm{i}\eta_{+}^i,\\[1mm]
\delta \zeta_{-}^{V i}=\,&\chi^{-1}\slashed{D}\chi\epsilon_{+}^{i}+2\mathrm{i}\eta_{-}^i.
\end{align}

We now consider the gauginos $\Omega_{\pm}^{Ii}$.  We take the linear combinations as in equation \eqref{Omega} and set their $Q$-variations to zero. We get,
\begin{subequations}\label{derivative-X}
\begin{align} 
&     \mathcal{D}_{0}(\chi^{-1/2}X_{+}^{I})=0,   
& \mathcal{D}_{p}(\chi^{-1/2}X_{+}^{I})=-h\chi^{-1/2}\left(F_{p0}^{I-}-\frac{1}{4}X_{-}^{I}T_{p0}^{-}\right)\label{F-}, \\
&    \mathcal{D}_{0}(\chi^{-1/2}X_{-}^{I})=0,
&   \mathcal{D}_{p}(\chi^{-1/2}X_{-}^{I})=h^{-1}\chi^{-1/2} \left(F_{p0}^{I+}-\frac{1}{4}X_{+}^{I}T_{p0}^{+}\right)\label{F+}, &
\end{align}
\end{subequations}
and
\be
Y^{ij \, I}=0.
\ee
From the background spinor ${\Psi}^i_{\pm}$ we get,
\begin{subequations} \label{derivative-A}
\begin{align}
      &  \mathcal{D}_{0}(\chi^{-1}{A}_{+})=0, &
        \mathcal{D}_{0}(\chi^{-1}{A}_{-}) &= 0 \\[1mm]
   & \mathcal{D}_{p}(\chi^{-1}{A}_{+})=-h {F}^{-}_{p0}\chi^{-1},&
      \mathcal{D}_{p}(\chi^{-1}{A}_{-})&=h^{-1} {F}^{+}_{p0}\chi^{-1}.
\end{align}
\end{subequations}
As in the full supersymmetry analysis of the previous section, we can now use the homogeneity property of $\mathcal{F}_I^{\pm}$ together with the definition of $G^{ab \pm}_{I}$, to obtain from \eqref{derivative-X} and \eqref{derivative-A} expressions for derivatives acting on  $\mathcal{F}_I^{\pm}$. We get,
\begin{align}
&     \mathcal{D}_{0}(\chi^{-1/2}\mathcal{F}_{I}^{+})= 0, \\  
 &     \mathcal{D}_{0}(\chi^{-1/2}\mathcal{F}_{I}^{-}) =0, \\
 &     \mathcal{D}_{p}(\chi^{-1/2}\mathcal{F}_{I}^{+})= h\chi^{-1/2}\left(G_{p0 I}^{-}-\frac{1}{4}\mathcal{F}^{-}_{I}T_{p0}^{-}\right),\label{G-}\\
 &       \mathcal{D}_{p}(\chi^{-1/2}\mathcal{F}_{I}^{-})= h^{-1}\chi^{-1/2}\left(G_{p0 I}^{+}+\frac{1}{4}\mathcal{F}^{+}_{I}T_{p0}^{+}\right)\label{G+}.
\end{align}

Now we turn to the variations of the Weyl multiplet fermions. We first consider the variation of $\chi_{\pm}^{i}$. We get,
\begin{align}
        \mathcal{D}_{c}(\chi^{1/2}T^{c0-})=\,&6 h D\, \chi^{1/2}\label{D},\\[1mm]
        \mathcal{D}_{c}(\chi^{1/2}T^{c0+})= \,& -6 h^{-1} D\, \chi^{1/2}\label{D1},\\[1mm]
    \mathcal{D}_{c}(\chi^{1/2}T^{cp-})=\,&-8h\chi^{1/2}R(A)^{p0-}\label{RA1},\\[1mm]
        \mathcal{D}_{c}(\chi^{1/2}T^{cp+})=\,&-8h^{-1}\chi^{1/2}R(A)^{p0+}\label{RA2},
\end{align}
and
\be
  R(\mathcal{V})_{ab}{}^{i}{}_{j\pm}=0. \label{RV0}
\ee
The variation of the gravitino field strength $R(Q)_{ab\pm}^{i}$ gives,
\begin{subequations} \label{tabcon}
\begin{align}
\mathcal{D}_{0}T_{ab}^{\pm}-\frac{1}{2}\chi^{-1}\mathcal{D}_{e}\chi\Big(\delta^{e}_{0}T_{ab}^{\pm}+2\delta_{0[a}T_{b]}{}^{e\pm}-2\delta^{e}_{[a}T_{b]0}^{\pm}\Big)=\,&0,\\
      \mathcal{D}_{p}T_{ab}^{-}-\frac{1}{2}\chi^{-1}\mathcal{D}_{e}\chi\Big(\delta^{e}_{p}T_{ab}^{-}+2\delta_{p[a}T_{b]}{}^{e-}-2\delta^{e}_{[a}T_{b]p}^{-}\Big)=\,& 8h \mathcal{R}(M)_{ab \,p0}^{-}\label{ModifiedRm-},\\
          \mathcal{D}_{p}T_{ab}^{+}-\frac{1}{2}\chi^{-1}\mathcal{D}_{e}\chi\Big(\delta^{e}_{p}T_{ab}^{+}+2\delta_{p[a}T_{b]}{}^{e +}-2\delta^{e}_{[a}T_{b]p}^{+}\Big)=\,& -8h^{-1} \mathcal{R}(M)_{ab \,p0}^{+}.\label{ModifiedRm+}
\end{align}
\end{subequations}

Now, we take the variation of the covariant derivative $D_{\mu}\bar{\zeta}_{i\pm}^{H}$, cf.~\eqref{derizeta}. Imposing the condition that $Q$-supersymmetry variation of the 
$D_{\mu}\Bar{\zeta}_{i+}^{H}$ 
vanishes, we get,
\begin{align}\label{zeta+}
 -\frac{1}{2}D_{\mu}(\chi^{-1}D^{a}\chi)  \,& +\frac{1}{4}(\chi^{-1}\mathcal{D}_{\mu}\chi)(\chi^{-1}\mathcal{D}^{a}\chi)-\frac{1}{8}(\chi^{-1}\mathcal{D}_{b}\chi)^2e_{\mu}^{a}\nonumber\\
   \,& -\frac{3}{4}(De_{\mu}{}^{a}-2De_{\mu}{}^{0}\delta^{a0})+2 R(A)^{+}_{\mu0}\delta^{0a}-(R(A)_{\mu}^{+}{}^{a}-R(A)_{\mu}^{-}{}^{a})=0. 
\end{align}
Similarly, from the variation of $D_{\mu}\Bar{\zeta}_{i-}^{H}$, we get,
\begin{align}\label{zeta-}
     -\frac{1}{2}D_{\mu}(\chi^{-1}D^{a}\chi)\,& +\frac{1}{4}(\chi^{-1}\mathcal{D}_{\mu}\chi)(\chi^{-1}\mathcal{D}^{a}\chi)-\frac{1}{8}(\chi^{-1}\mathcal{D}_{b}\chi)^2e_{\mu}^{a}\nonumber\\
   \,& -\frac{3}{4}(De_{\mu}{}^{a}-2De_{\mu}{}^{0}\delta^{a0})-2 R(A)^{-}_{\mu0}\delta^{0a}-(R(A)_{\mu}^{+}{}^{a}-R(A)_{\mu}^{-}{}^{a})=0.
\end{align}
Subtracting \eqref{zeta+} from \eqref{zeta-} gives,
\begin{align}
    R(A)_{a0}=\widetilde{R}(A)_{pq}=0,
\end{align}
and adding \eqref{zeta+} and \eqref{zeta-} gives,
\begin{align}
   \,& -D_{\mu}(\chi^{-1}D^{a}\chi)   +\frac{1}{2}(\chi^{-1}\mathcal{D}_{\mu}\chi)(\chi^{-1}\mathcal{D}^{a}\chi)-\frac{1}{4}(\chi^{-1}\mathcal{D}_{b}\chi)^2e_{\mu}^{a}\nonumber\\
   \,& -\frac{3}{2} \left(De_{\mu}{}^{a}-2De_{\mu}{}^{0}\delta^{a0}\right)+2(R(A)^{+}_{\mu0}\delta^{0a}-R(A)^{-}_{\mu0}\delta^{0a})-2(R(A)_{\mu}^{+}{}^{a}-R(A)_{\mu}^{-}{}^{a})=0.
\end{align}
From this equation, we can extract the gauge field $f_{\mu}{}^{a}$ by expanding the supercovariant derivatives. We find,
\begin{align}
    f_{\mu}{}^{a}= \,& -\frac{1}{2}\mathcal{D}_{\mu}(\chi^{-1}\mathcal{D}^{a}\chi)   +\frac{1}{4}(\chi^{-1}\mathcal{D}_{\mu}\chi)(\chi^{-1}\mathcal{D}^{a}\chi)-\frac{1}{8}(\chi^{-1}\mathcal{D}_{b}\chi)^2e_{\mu}^{a}\nonumber\\
   \,& -\frac{3}{4}(De_{\mu}{}^{a}-2De_{\mu}{}^{0}\delta^{a0})+\widetilde{R}(A)_{\mu0}\delta^{0a}-\widetilde{R}(A)_{\mu}{}^{a}.
\end{align}
On the other hand using the explicit expression of $f_{\mu}{}^{a}$ given in \eqref{fmua}, we can conclude
\begin{eqnarray}
\label{ricciriem}
R(\omega,e)_{\mu}{}^{a} & - & \frac{1}{6}R(\omega, e)e_{\mu}{}^{a} \nonumber \\
          &=&  -\mathcal{D}_{\mu}(\chi^{-1}\mathcal{D}^{a}\chi)   +\frac{1}{2}(\chi^{-1}\mathcal{D}_{\mu}\chi)(\chi^{-1}\mathcal{D}^{a}\chi)-\frac{1}{4}(\chi^{-1}\mathcal{D}_{b}\chi)^2e_{\mu}^{a}\nonumber\\
   &&  -(De_{\mu}{}^{a}-3De_{\mu}{}^{0}\delta^{a0})+\frac{1}{16}T_{\mu b}^{-}T^{+ba} +(\widetilde{R}(A)_{\mu0}\delta^{0a}-\widetilde{R}(A)_{0}{}^{a}e_{\mu}^{0}).
        \end{eqnarray}

The constraints obtained so far ensure that the $Q$-variation of the background spinor $\Lambda^{i}$ vanishes.

We now examine the implications of the gauge conditions that 
eliminates the freedom of making local scale transformations and conformal boosts. This gauge conditions amount to choosing $b_\mu = 0$ and $\chi =$ constant. Using the gauge condition $\chi=\text{constant}$ in \eqref{ricciriem}, we find that the Ricci scalar is proportional to the auxiliary field $D$,
\begin{align}\label{Dequalricci}
    R(\omega,e)=-3D.
\end{align}
Using this fact in equations \eqref{D}--\eqref{RA2}, we obtain
\begin{align}
&    h^{-1}\mathcal{D}^{p}T_{p0}^{-}= 6D=-h\mathcal{D}^{p}T_{p0}^{+}\label{DFinal},\\[1mm]
 &   h \, \mathcal{D}_{[p}T_{q]0}^{+}= 2R(A)_{pq}=h^{-1}\mathcal{D}_{[p}T_{q]0}^{-},\label{RAfinal}
    \end{align}
and in equations \eqref{tabcon} we get,
\begin{subequations}
    \begin{align}
    \mathcal{R}(M)_{pq \, 0l}=\,&-\frac{1}{8}h^{-1}\varepsilon_{pq}{}^{s}\mathcal{D}_{l}T_{s0}^{-} -\frac{1}{8}h\,\varepsilon_{pq}{}^{s}\mathcal{D}_{l}T_{s0}^{+},\\[1mm]
    \mathcal{R}(M)_{0p\, 0q}=\,&\frac{1}{8}h^{-1}\mathcal{D}_{q}T_{p0}^{-}-\frac{1}{8}h\mathcal{D}_{q}T_{p0}^{+},\\[1mm] 
    \mathcal{R}(M)_{0l\,pq}=\,&-\frac{1}{8}h^{-1}\varepsilon_{pq}{}^{s}\mathcal{D}_{s}T_{l0}^{-}-\frac{1}{8}h\varepsilon_{pq}{}^{s}\mathcal{D}_{s}T_{l0}^{+},\\[1mm]
    \mathcal{R}(M)_{pq\,ls}=\,&\frac{1}{8}\varepsilon_{ls}{}^{v}\varepsilon_{pq}{}^{u}h^{-1}\mathcal{D}_{v}T_{u0}^{-}-\frac{1}{8}\varepsilon_{ls}{}^{v}\varepsilon_{pq}{}^{u}h\mathcal{D}_{v}T_{u0}^{+}.
\end{align}
\end{subequations}
We can now express the components of the Riemann tensor in terms of the $T$-fields as, 
\begin{subequations}
\label{R1}
\begin{align}
    R(\omega)_{pq}{~}_{0r}&=\frac{1}{8}\varepsilon_{pq}{}^s\left(-h^{-1}\mathcal{D}_r T^-_{s0}-h\mathcal{D}_r T^+_{s0} - \frac{1}{4}\left(T^-_{s0}T^+_{0r}-T^+_{s0}T^-_{0r} \right)  \right),  \\[1mm]
     R(\omega)_{0p}{~}_{0q}&=\frac{1}{8}\left(h^{-1}\mathcal{D}_{q}T_{p0}^{-}-h\mathcal{D}_{q}T_{p0}^{+} - \frac{1}{4}\left( T^-_{0p}T^+_{0q}+T^+_{0p}T^-_{0q}  \right)\right),\\[1mm]
    R(\omega)_{pq}{~}_{rs}&=-\frac{1}{2}\delta_{[r[p}\left(h^{-1}\mathcal{D}_{q]}T_{s]0}^{-}-h\mathcal{D}_{q]}T_{s]0}^{+}\right)\nonumber\\
    \,&+\frac{1}{8}\left(\delta_{[r[p}\delta_{q]s]}T^{-u0}T_{u0}^{+}+T_{[p0}^{+}T_{0[r}^{-}\delta_{q]s]}+T_{[p0}^{-}T_{0[r}^{+}\delta_{q]s]}\right).
\end{align}
\end{subequations}
Now we choose vielbein components appropriately---a gauge condition for local rotational invariance. The choice 
\be
e_{\tau}{}^{p}=0
\ee
is consistent with the embedding condition \eqref{condembedd} which breaks the four-dimensional rotational covariance but has manifest three-dimensional rotational covariance.  We also impose that nothing depends on the Euclidean time $\tau$. Denoting the four-dimensional world indices as $(\tau, m)$, we parameterize the vielbein as,
\begin{align}
    e_{\mu}{}^{0}dx^{\mu}=\,&e^g (d\tau+ \sigma_{m}dx^{m}),\\
    e_{\mu}{}^{p}dx^{\mu}=\,&e^{-g}\hat{e}_{m}{}^{p}dx^{m},
\end{align}
here $\hat{e}_{m}{}^{p}$ is the rescaled three-dimensional vielbein of the three-dimensional space. The components of the inverse vielbein $(E_{a}{}^{\mu})$ are,
\begin{align}
    E_{0}{}^{\tau}=e^{-g},~~E_{0}{}^{m}=0,~~E_{p}{}^{\tau}=-\sigma_{p}e^{g},~~E_{p}{}^{m}=e^{g}\hat{E}_{p}{}^{m}.
\end{align}
The spin connection coefficients read
\begin{align}
\omega_{l \, pq}=\,&e^{g}\left(\hat{\omega}_{l\,pq}+2\delta_{l[p}\nabla_{q]}g\right), &
    \omega_{0\, 0p}=\,&-e^{g}\nabla_{p}g, \\
    \omega_{0 \, pq}=\,&\frac{1}{2}e^{3g}\varepsilon_{pql}R(\sigma)^{l}, &
\omega_{p\, q0}=\,&-\frac{1}{2}e^{3g}\varepsilon_{pql}R(\sigma)^{l},
\end{align}
where $R(\sigma)^{l}=\varepsilon^{lpq}\nabla_{p}\sigma_{q}$ and where $\hat{\omega}_{l\, pq}$ are the three-dimensional spin connection coefficients and 
\begin{align}
\nabla_{p}=\hat{E}_{p}{}^{m}\nabla_{m}   
\end{align}
In the equations below, we will use these expressions as and when needed. The corresponding curvature components are (where we use three-dimensional notation on the right-hand side), 
\begin{subequations}\label{R}
 \begin{align}
R(\omega)_{pq \, 0r}=& \,  -\frac{1}{2}\varepsilon_{pq}{}^s e^{4g}\left( \nabla_r R(\sigma)_s+5 R(\sigma)_s\nabla_rg+R(\sigma)_r\nabla_sg-2\delta_{sr}R(\sigma)^l\nabla_lg 
\right),\\
R(\omega)_{0p \, 0q}=\, & \, e^{2g}\left(\nabla_p\nabla_qg+3\nabla_pg \nabla_qg-\delta_{pq}\left(\nabla_sg\right)^2\right)+\frac{1}{4}e^{6g}\left(R(\sigma)_pR(\sigma)_q 
-\delta_{pq}\left(R(\sigma)\right)^2\right),\\
R(\omega)_{pq \, rs}=\, & \, e^{2g}\widehat{R}_{pq \, rs}+3 e^{6g}\delta_{[p[r}\Big(\frac{1}{2}\delta_{s]q]}R(\sigma)_{l}R(\sigma)^{l}-R(\sigma)_{s]}R(\sigma)_{q]}\Big)\nonumber\\
\,&-4e^{2g}\delta_{[p[r}\Big(\nabla_{s]}\nabla_{q]}g+ \nabla_{s]}g\nabla_{q]}g-\frac{1}{2}\delta_{s]q]}\nabla^l g\nabla_{l}g\Big)
    \end{align}   
\end{subequations}
where $\widehat{R}_{pq \, rs}$ are the components of the Riemann tensor of the three-dimensional base space.

So far, we have only considered variations of covariant quantities with respect to the full ${\cal N} =2$ supersymmetry. However, under the residual supersymmetry, the linear combination 
\be
\psi_{\mu+}^{i}- \mathrm{i} h\gamma_{0}\psi_{\mu-}^{i}
\ee 
transforms covariantly. Hence, we  get new conditions by setting its $Q$-supersymmetry variation to zero.

The transformation of the chiral gravitino evaluated in the half supersymmetric background takes the form,  
\begin{align}
    \delta \psi_{\tau+}^{i}=\,&2\partial_{\tau}\epsilon_{+}^{i}+A_{\tau}\epsilon_{+}^{i}+e^{2g}\left(T_{p}^{-}-\nabla_{p}g-\frac{1}{2}e^{2g}R(\sigma)_{p}\right)\gamma^{p}\gamma_{0}\epsilon_{+}^{i},\\
    \delta_{}\psi_{m+}^{i}=\,&2\nabla_{m}\epsilon_{+}^{i}-(T_{m}^{-}-A_{m})\epsilon_{+}^{i}\nonumber\\
    \,&-\hat{e}_{m}^{p}\varepsilon_{p}{}^{qr}\left(T_{r}^{-}-\nabla_{r}g-\frac{1}{2}e^{2g}R(\sigma)_{r}\right)\gamma_{q}\gamma_{0}\epsilon_{+}^{i}\nonumber\\
    \,& +e^{2g}\sigma_{m}\left(T_{p}^{-}-\nabla_{p}g-\frac{1}{2}e^{2g}R(\sigma)_{p}\right)\gamma^{p}\gamma_{0}\epsilon_{+}^{i},
\end{align}
where we have defined the three dimensional world vectors $T_{m}^{+}$ and $T_{m}^{-}$ as,
    \begin{align}
T_{m}^{+}\equiv\,&\frac{1}{4}he^{-g}\hat{e}_{m}{}^{p}T_{p0}^{+},~T_{p}^{+}=\hat{E}_{p}{}^{m}T_{m}^{+}\label{Tm+}, \\
T_{m}^{-}\equiv\,&\frac{1}{4}h^{-1}e^{-g}\hat{e}_{m}{}^{p}T_{p0}^{-},~T_{p}^{-}=\hat{E}_{p}{}^{m}T_{m}^{-}\label{Tm-}.
    \end{align}
The transformation of the anti-chiral gravitino takes the form, 
\begin{align}
    \delta \psi_{\tau-}^{i}=\,&2\partial_{\tau}\epsilon_{-}^{i}-A_{\tau}\epsilon_{-}^{i}-e^{2g}\left(T_{p}^{+}+\nabla_{p}g-\frac{1}{2}e^{2g}R(\sigma)_{p}\right)\gamma^{p}\gamma_{0}\epsilon_{-}^{i},\\
    \delta_{}\psi_{m-}^{i}=\,&2\nabla_{m}\epsilon_{-}^{i}+(T_{m}^{+}-A_{m})\epsilon_{-}^{i}\nonumber\\
    \,&-\hat{e}_{m}{}^{p}\varepsilon_{p}{}^{qr}\left(T_{r}^{+}+\nabla_{r}g-\frac{1}{2}e^{2g}R(\sigma)_{r}\right)\gamma_{q}\gamma_{0}\epsilon_{-}^{i}\nonumber\\
    \,& -e^{2g}\sigma_{m}\left(T_{p}^{+}+\nabla_{p}g-\frac{1}{2}e^{2g}R(\sigma)_{p}\right)\gamma^{p}\gamma_{0}\epsilon_{-}^{i}.
\end{align}
Demanding the variation $\delta\psi_{\mu+}^{i}-\mathrm{i} h\gamma_{0}\delta\psi_{\mu-}^{i}$  to vanish under $Q$-supersymmetry yields the following conditions 
    \begin{align}
&        T_{p}^{-}=\nabla_{p}g+\frac{1}{2}e^{2g}R(\sigma)_{p}\label{Tp-},\\
&  T_{p}^{+}=-\nabla_{p}g+\frac{1}{2}e^{2g}R(\sigma)_{p}\label{Tp+},\\
&  h^{-1}\nabla_{m}h+A_{m}=\frac{1}{2}e^{2g}R(\sigma)_{m}\label{h}.
    \end{align}
These  are the key  results for the rest of the analysis. With these results we return to the previous identities. 

To start with, using definitions \eqref{Tm+} and \eqref{Tm-} we can write, 
\begin{align}
& h^{-1}\mathcal{D}_{p}T_{q0}^{-}= 4e^{2g}\left(\nabla_{p}T_{q}^{-}+2T_{p}^{-}T_{q}^{-}-\delta_{pq}T_{s}^{-}T^{s-}\right),\\[1mm]
& h \, \mathcal{D}_{p}T_{q0}^{+}= 4e^{2g}\left(\nabla_{p}T_{q}^{+}-2T_{p}^{+}T_{q}^{+}+\delta_{pq}T_{s}^{+}T^{s+}\right),
\end{align}
where note that the right hand is expressed completely in terms of the three dimensional vector $T_{m}^{\pm}$.
Inserting these expressions in equations \eqref{DFinal} and \eqref{RAfinal}, we obtain
\begin{align}\label{Dterm}
    & D=\frac{2}{3}e^{2g}\left[\nabla_{p}^2g-\nabla_{p}g\nabla^{p}g-\frac{1}{4}e^{4g}R(\sigma)_{p}R(\sigma)^{p}\right],\\
    & R(A)_{pq}= e^{4g}\left[2\nabla_{[p}g R(\sigma)_{q]}+\nabla_{[p}R(\sigma)_{q]}\right].
\end{align}
We note that this is consistent with \eqref{Dequalricci}.
Additionally, it turns out that \eqref{R1} and \eqref{R} agree as long as the three-dimensional Riemann curvature is zero,
\be
\widehat{R}_{pq \, rs } =0.
\ee
We conclude that  the three-dimensional base space is flat.

Next we consider components of the Maxwell gauge field strengths. They take more complicated form compared to their fully supersymmetric counterparts. Using 
\eqref{F-} and \eqref{F+} we obtain,
\begin{align}
    F_{p0}^{-I}=\,&e^{g}\Big[-\nabla_{p}(h^{-1}X_{+}^{I})+(\nabla_{p}g)h X_{-}^{I} -\frac{1}{2}e^{2g}R(\sigma)_{p}(h^{-1}X_{+}^{I}-h X_{-}^{I})\Big]\label{Fp0+},\\
      F_{p0}^{+I}=\,&e^{g}\Big[\nabla_{p}(h X_{-}^{I})-(\nabla_{p}g)h^{-1} X_{+}^{I}+\frac{1}{2}e^{2g}R(\sigma)_{p}(h^{-1}X_{+}^{I}-h X_{-}^{I})\Big].\label{Fp0-}
\end{align}
The dual field strengths can be obtained from \eqref{G-} and \eqref{G+} in a similar fashion. We get,  
\begin{align}
    G_{p0 I}^{-}=\,&e^{g}\Big[\nabla_{p}(h^{-1}\mathcal{F}_{I}^{+})+(\nabla_{p}g)h \mathcal{F}_{I}^{-}+\frac{1}{2}e^{2g}R(\sigma)_{p}(h\mathcal{F}_{I}^{-}+h^{-1}\mathcal{F}_{I}^{+})\Big],\\
    G_{p0 I}^{+}=\,&e^{g}\Big[\nabla_{p}(h\mathcal{F}_{I}^{-})+(\nabla_{p}g)h^{-1} \mathcal{F}_{I}^{+}-\frac{1}{2}e^{2g}R(\sigma)_{p}(h\mathcal{F}_{I}^{-}+h^{-1}\mathcal{F}_{I}^{+})\Big].
\end{align}
We can now analyze the Maxwell equations. First, we derive a series of expressions for the derivatives of the gauge field strengths,
\begin{align}
\mathcal{D}^{a}F_{ap}^{I-}=\,&\varepsilon_{p}{}^{qr}e^g\Bigg[\frac{1}{2}\nabla_{q}\Big(e^{3g}R(\sigma)_{r}(h^{-1}X_{+}^{I}-hX_{-}^{I})\Big)+e^{g}\nabla_{q}g\nabla_{r}(h^{-1}X_{+}^{I}+hX_{-}^{I})\Bigg],\\
\mathcal{D}^{a}F_{ap}^{I+}=\,&\varepsilon_{p}{}^{qr}e^g\Bigg[\frac{1}{2}\nabla_{q}\Big(e^{3g}R(\sigma)_{r}(h^{-1}X_{+}^{I}-hX_{-}^{I})\Big)+e^{g}\nabla_{q}g\nabla_{r}(h^{-1}X_{+}^{I}+hX_{-}^{I})\Bigg],\\
\mathcal{D}^{p}F_{p0}^{I-}=\,&e^{2g}\Bigg[\frac{1}{2}e^{4g}R(\sigma)_{p}R(\sigma)^{p}(h^{-1}X_{+}^{I}-h X_{-}^{I})+\frac{1}{2}e^{3g}R(\sigma)_{p}\nabla^p\Big[e^{-g}(h X_{-}^{I}+h^{-1}X_{+}^{I})\Big]\nonumber\\
\,&-\nabla_{p}^{2}(h^{-1}X_{+}^{I})+(\nabla_{p}^{2}g) h X_{-}^{I}+\nabla^{p}g \nabla_{p}(h^{-1}X_{+}^{I}+hX_{-}^{I})-(\nabla^{p}g \nabla_{p}g) h X_{-}^{I}\Bigg],\\
\mathcal{D}^{p}F_{p0}^{I+}=\,&e^{2g}\Bigg[\frac{1}{2}e^{4g}R(\sigma)_{p}R(\sigma)^{p}(h^{-1}X_{+}^{I}-h X_{-}^{I})+\frac{1}{2}e^{3g}R(\sigma)_{p}\nabla^p\Big[e^{-g}(h X_{-}^{I}+h^{-1}X_{+}^{I})\Big]\nonumber\\
\,&+\nabla_{p}^{2}(h X_{-}^{I})-(\nabla_{p}^{2}g) h^{-1} X_{+}^{I}-\nabla^{p}g \nabla_{p}(h^{-1}X_{+}^{I}+hX_{-}^{I})+(\nabla^{p}g \nabla_{p}g) h^{-1}X_{+}^{I}\Bigg],\\
\mathcal{D}^{a}G_{apI}^{-}=\,&-e^g\varepsilon_{p}{}^{qr}\Bigg[\frac{1}{2}\nabla_{q}\Big(e^{3g}R(\sigma)_{r}(h\mathcal{F}_{I}^{-}+h^{-1}\mathcal{F}_{I}^{+})\Big)-e^{g}\nabla_{q}g\nabla_{r}(h\mathcal{F}_{I}^{-}-h^{-1}\mathcal{F}_{I}^{+})\Bigg],\\
\mathcal{D}^{a}G_{apI}^{+}=\,&-e^g\varepsilon_{p}{}^{qr}\Bigg[\frac{1}{2}\nabla_{q}\Big(e^{3g}R(\sigma)_{r}(h\mathcal{F}_{I}^{-}+h^{-1}\mathcal{F}_{I}^{+})\Big)-e^{g}\nabla_{q}g\nabla_{r}(h\mathcal{F}_{I}^{-}-h^{-1}\mathcal{F}_{I}^{+})\Bigg],\\
\mathcal{D}^{q}G_{q0I}^{-}=\,&e^{2g}\Bigg[\nabla_{p}^2(h^{-1}\mathcal{F}_{I}^{+})+(\nabla_{p}^2 g) h \mathcal{F}_{I}^{-}+\nabla_{p}g\nabla^{p}(h\mathcal{F}_{I}^{-}-h^{-1}\mathcal{F}_{I}^{+})-(\nabla^{p}g\nabla_{p}g) h\mathcal{F}_{I}^{-}\nonumber\\
\,&+\frac{1}{2}e^{3g}R(\sigma)_{p}\nabla^{p}\Big[e^{-g}(h\mathcal{F}_{I}^{-}-h^{-1}\mathcal{F}_{I}^{+})\Big] -\frac{1}{2}e^{4g}(R(\sigma)_{p})^2(h\mathcal{F}_{I}^{-}+h^{-1}\mathcal{F}_{I}^{+}
)\Bigg],\\
\mathcal{D}^{q}G_{q0I}^{+}=\,&e^{2g}\Bigg[\nabla_{p}^2(h\mathcal{F}_{I}^{-})+(\nabla_{p}^2 g) h^{-1}\mathcal{F}_{I}^{+}-\nabla_{p}g\nabla^{p}(h\mathcal{F}_{I}^{-}-h^{-1}\mathcal{F}_{I}^{+})-(\nabla^{p}g\nabla_{p}g) h^{-1}\mathcal{F}_{I}^{+}\nonumber\\
\,&+\frac{1}{2}e^{3g}R(\sigma)_{p}\nabla^{p}\Big[e^{-g}(h\mathcal{F}_{I}^{-}-h^{-1}\mathcal{F}_{I}^{+})\Big] -\frac{1}{2}e^{4g}(R(\sigma)_{p})^2(h\mathcal{F}_{I}^{-}+h^{-1}\mathcal{F}_{I}^{+}
)\Bigg].
\end{align}
Using these expressions in  \eqref{eom} and \eqref{bian}, we obtain two non-trivial Maxwell equations,
\begin{align}
\nabla_{p}^{2}\Big[e^{-g}\Big(h X_{-}^{I}+h^{-1}X_{+}^{I}\Big)\Big]=\,&0, \label{harmonic1}\\
\nabla_{p}^{2}\Big[e^{-g}\Big(h \mathcal{F}^{-}_{I}-h^{-1}\mathcal{F}^{+}_{I}\Big)\Big]=\,&0. \label{harmonic2}
\end{align}

Until this point, our analysis holds for an arbitrary chiral background. Now we identify the (anti-) chiral background as 
\begin{align}\label{Apm}
A_{+}=& 64e^{2g}h^2(T_{p}^{-})^2,\\[1mm] 
A_{-}=& 64e^{2g}h^{-2}(T_{p}^{+})^2,
\end{align}
with $T_p^\pm$ obtained above. Putting all the constraints in the $D$ equation of motion \eqref{Deom} we  get, 
    \begin{align} \label{final-g-eq}
    e^{-\mathcal{K}}+\frac{1}{2}\chi=\,&-128e^{3g}\nabla_{p}\Big[e^{-g}\nabla^{p}g(\mathcal{F}_{A}^{+}+\mathcal{F}_{A}^{-})\Big]+32 e^{6g}R(\sigma)_{p}^{2}(\mathcal{F}_{A}^{+}+\mathcal{F}_{A}^{-})\nonumber\\
    \,&-64e^{4g}R(\sigma)_{p}
\nabla^{p}(\mathcal{F}_{A}^{+}-\mathcal{F}_{A}^{-}).
\end{align}
Putting all the constraints in $A_{\mu}$ equation \eqref{EOMAMU} we get  after a tedious algebra, 
\begin{align} \label{final-sigma-eq}
  -(h^{-1}X_{+}^{I}+\,&hX_{-}^{I})\overset{\leftrightarrow}{\nabla}_{p}(h\mathcal{F}_{I}^{-}-h^{-1}\mathcal{F}_{I}^{+})-\frac{1}{2}\chi e^{2g}R(\sigma)_{p}\nonumber\\
=\,&128 e^{2g}\nabla^{q}\Bigg[2\nabla_{[q}g\nabla_{p]}\Big(\mathcal{F}_{A}^{+}-\mathcal{F}_{A}^{-}\Big)-\nabla_{[q}\Big(e^{2g}R(\sigma)_{p]}(\mathcal{F}_{A}^{+}+\mathcal{F}_{A}^{-})\Big)\Bigg].
\end{align}
The above two equations determine $g$ and $\sigma_m$.
This concludes our analysis. The solutions can now be expressed in terms of harmonic functions according to \eqref{harmonic1}--\eqref{harmonic2}. The two field equations \eqref{final-g-eq} and \eqref{final-sigma-eq} then determine the function $g$ and one-form $\sigma_m$, from which the full solution can be constructed. Equations \eqref{harmonic1}, \eqref{harmonic2}, \eqref{final-g-eq} and \eqref{final-sigma-eq} are the key results of this paper.

We end this section by writing equations \eqref{harmonic1}--\eqref{harmonic2} in terms of rescaled variables \cite{Behrndt:1996jn, LopesCardoso:2000qm},
\begin{align}
    &Y_{+}^{I}=e^{-g}h^{-1}X_{+}^{I}, & &\Upsilon_{+}=e^{-2g}h^{-2}A_{+},\\
    &Y_{-}^{I}=e^{-g}h X_{-}^{I}, & & \Upsilon_{-}=e^{-2g}h^{2}A_{-}.
\end{align}
In these variables, the Maxwell equations read
\begin{align}
\nabla_{p}^{2}\Big[Y_{-}^{I}+Y_{+}^{I}\Big]=0,\\
\nabla_{p}^{2}\Big[\mathcal{F}^{-}_{I}(Y_{-}, \Upsilon_{-})-\mathcal{F}^{+}_{I}(Y_{+}, \Upsilon_{+})\Big]=0, 
\end{align}
which imply,
\begin{align}
    Y_{-}^{I}+Y_{+}^{I}=H^{I}, \label{gse}\\
    \mathcal{F}^{-}_{I}(Y_{-}, \Upsilon_{-})-\mathcal{F}^{+}_{I}(Y_{+}, \Upsilon_{+})=H_{I},
\end{align}
where $H^{I}$ and $H_{I}$ are harmonic functions in the three-dimensional base space. These are the Euclidean version of the generalised stabilisation equations.

\section{Summary of the results for the half-BPS analysis}\label{sec:summary}
For the convenience of the reader, this section collects the essential results and key equations required to construct half-BPS solutions.\footnote{This section has been added at the suggestion of the referee.} In particular, we highlight that all bosonic fields of the half-BPS configurations are completely determined in terms of harmonic functions on the three-dimensional base space.

The general form of the half-BPS metric is
\begin{align}
    ds^2=e^{2g}(d\tau+\sigma_{m}dx^{m})^2+e^{-2g}dx_{m}dx^{m},
\end{align}
where the three-dimensional base space is flat. The three-dimensional Hodge dual of the two-form $d\sigma$ is defined as, 
\begin{align}
    R(\sigma)^{p}=\varepsilon^{pql}\nabla_{q}\sigma_{l},
\end{align}
and will appear repeatedly below.

The auxiliary fields are determined by the following equations
\begin{align}
      D=\,&\frac{2}{3}e^{2g}\left[\nabla_{p}^2g-\nabla_{p}g\nabla^{p}g-\frac{1}{4}e^{4g}R(\sigma)_{p}R(\sigma)^{p}\right],\\
     R(A)_{pq}= \,&e^{4g}\left[2\nabla_{[p}g R(\sigma)_{q]}+\nabla_{[p}R(\sigma)_{q]}\right],\\
    T_{p0}^{+}=\,&4h^{-1}e^gT_{p}^{+},~~~ T_{p}^{+}=-\nabla_{p}g+\frac{1}{2}e^{2g}R(\sigma)_{p},\\
     T_{p0}^{-}=\,&4he^gT_{p}^{-},~~~ T_{p}^{-}=\nabla_{p}g+\frac{1}{2}e^{2g}R(\sigma)_{p},\\
     R(\mathcal{V})_{ab}{}^{i}{}_{j}=\,&0,\\
     Y^{I~ij}=\,&0.
\end{align}
Substituting the auxiliary fields obtained above into the 
$D$-equation of motion, cf.~\eqref{Deom}, yields the differential equation determining the metric function $g$:
    \begin{align} 
    e^{-\mathcal{K}}+\frac{1}{2}\chi=\,&-128e^{3g}\nabla_{p}\Big[e^{-g}\nabla^{p}g(\mathcal{F}_{A}^{+}+\mathcal{F}_{A}^{-})\Big]+32 e^{6g}R(\sigma)_{p}^{2}(\mathcal{F}_{A}^{+}+\mathcal{F}_{A}^{-})\nonumber\\
    \,&-64e^{4g}R(\sigma)_{p}
\nabla^{p}(\mathcal{F}_{A}^{+}-\mathcal{F}_{A}^{-}).
\end{align}
Substituting the auxiliary fields obtained above into the $A_\mu$-equations of motion, cf. \eqref{EOMAMU}, yields the differential equation determining the three-dimensional one-form $\sigma_m$
    \begin{align}
   -(h^{-1}X_{+}^{I}+\,&hX_{-}^{I})\overset{\leftrightarrow}{\nabla}_{p}(h\mathcal{F}_{I}^{-}-h^{-1}\mathcal{F}_{I}^{+})-\frac{1}{2}\chi e^{2g}R(\sigma)_{p}\nonumber\\
=\,&128 e^{2g}\nabla^{q}\Bigg[2\nabla_{[q}g\nabla_{p]}\Big(\mathcal{F}_{A}^{+}-\mathcal{F}_{A}^{-}\Big)-\nabla_{[q}\Big(e^{2g}R(\sigma)_{p]}(\mathcal{F}_{A}^{+}+\mathcal{F}_{A}^{-})\Big)\Bigg].
\end{align}
Finally, the scalar fields are determined in terms of harmonic functions through the generalised stabilization equations:
\begin{align}
    Y_{-}^{I}+Y_{+}^{I}=H^{I}, \label{gse1}\\
    \mathcal{F}^{-}_{I}(Y_{-}, \Upsilon_{-})-\mathcal{F}^{+}_{I}(Y_{+}, \Upsilon_{+})=H_{I},
\end{align}
where the rescaled variables are defined as 
\begin{align}
    &Y_{+}^{I}=e^{-g}h^{-1}X_{+}^{I}, & &\Upsilon_{+}=e^{-2g}h^{-2}A_{+},\\
    &Y_{-}^{I}=e^{-g}h X_{-}^{I}, & & \Upsilon_{-}=e^{-2g}h^{2}A_{-}.
\end{align}

At this stage, it is useful to restrict attention to the two-derivative theory and to count the number of variables to be solved for against the number of available equations. The unknowns consist of $(2n+2)$ real scalars, the functions $h$ and $g$, and the three components of $R(\sigma)_{p}$.  For two derivative theories the $D$- and $A_{\mu}$-equations take the form 
\begin{align}
    e^{-\mathcal{K}}+\frac{1}{2}\chi=\,&0, \label{simple-D-equation}\\
    H^{I}\overset{\leftrightarrow}{\nabla}_{p}H_{I}=\,&-\frac{1}{2}\chi R(\sigma)_{p}.
\end{align}
Imposing the dilatation gauge-fixing condition $\chi=-2$ in \eqref{simple-D-equation} yields,
\begin{align}
    e^{-\mathcal{K}}=\mathcal{F}^{+}_{I} X_{-}^{I}+X_{+}^{I}\mathcal{F}^{-}_{I}=1. \label{e-K}
\end{align}
Earlier,  we noted that the $D$-equation may be regarded as determining the function $g$. Strictly speaking, this interpretation depends on the choice of gauge. In the two-derivative theory, it is more natural to view
equation~\eqref{e-K}  as imposing a constraint on one of the $(2n+2)$ real scalars, which we choose to be $X_{+}^{0}$,  in terms of the remaining $2n+1$ scalars.  In addition, the $SO(1,1)$ chiral symmetry must be fixed, which we accomplish by imposing
\begin{align}
    X_{+}^{0}=-X_{-}^{0},
\end{align}
which leaves $2n$ real scalars to be determined. The complete set of $(2n+2)$ variables --- namely the 
$2n$ scalars together with the functions 
$g$ and 
$h$ --- is fixed by the $(2n+2)$ stabilization equations. Finally, the components of $R(\sigma)_{p}$ can be determined in terms of the harmonic functions $H^{I}, H_{I}$.

\section{Conclusions}\label{conclusion}
In this paper, we have analyzed fully BPS and a broad class of half-BPS stationary solutions of higher-derivative $\mathcal{N}=2$ Euclidean supergravity.  In section \ref{sec:Full}, we showed that Euclidean $AdS_{2}\times S^{2}$ is a fully BPS solution to the theory. We derived the Euclidean attractor equations and evaluated the Wald entropy for Euclidean $AdS_{2}\times S^{2}$. We also discussed a novel class of configurations where the metric is flat but other fields are non-trivial. We believe that these configurations are crucial to understanding the new attractor mechanism in Euclidean supergravity, which we plan to explore in our future work.

In section \ref{sec:half}, we studied  half-BPS solutions. We restricted our analysis  to the class of solutions that satisfy the embedding condition \eqref{condembedd}.
We derived the generalized stabilization equations that express the half-BPS solutions in terms of harmonic functions. Although, naively some minus signs look out of place, our final results are all  covariant with respect to electric-magnetic duality, wherein
\be
 \begin{pmatrix} 
     X_\pm^I \\[1mm]
      \mp\mathcal{F}^\pm_I
    \end{pmatrix}
    \ee
form symplectic pairs~\cite{deWit:2017cle}.  The fact that our final results---equations
\eqref{harmonic1}, \eqref{harmonic2}, \eqref{final-g-eq} and \eqref{final-sigma-eq}---are similar to the corresponding Lorentzian results \cite{LopesCardoso:2000qm} confirms that a large class of physically interesting half-BPS solutions, e.g., those obtained by analytic continuation, satisfy our embedding condition \eqref{condembedd}.

In \cite{Ciceri:2023mjl}, a detailed dictionary between Euclidean and Lorentzian supergravity was obtained for fermions and gauge fields, but not for the scalar sector. It therefore remains unclear how to relate the Lorentzian scalars and the associated prepotential to their Euclidean counterparts. This subtlety originates from the fact that the Euclidean scalar manifold is special para-K\"ahler, whereas the Lorentzian theory is based on special K\"ahler geometry. As a result, the two theories are not related by a straightforward Wick rotation. In particular, the precise relation between the Euclidean entropy formula derived here and its Lorentzian counterpart, as well as between the Euclidean and Lorentzian half-BPS equations, remains unclear. Constructing a detailed Lorentzian–Euclidean dictionary for scalar fields is therefore an important direction for future work.

As mentioned in the introduction, in the higher derivative $\mathcal{N}=2$ supergravity, although we know almost nothing in detail about the saddle solutions (for the path integral approach to computing the gravitational index), through a detailed analysis of the equations of motion it is possible to show that the gravitational index equals the Wald entropy \cite{Hegde:2024bmb}. This relation was conjectured in \cite{Chen:2024gmc}.  The matching shown in \cite{Hegde:2024bmb} requires certain analytic continuation. The analytic continuation rules are most certainly well motivated, but we would like to understand the matching purely in the Euclidean signature. One of the main motivations of this work was to make progress on this question. We have taken several steps in that direction. We have obtained the Wald entropy for a general extremal single-centre half-BPS black hole in the Euclidean supergravity notation. We have obtained the equations of motion satisfied by the half-BPS solutions. These are two main technical inputs that go into the analysis of \cite{Chen:2024gmc, Hegde:2024bmb}.  In order to complete the program, we need to understand the new attractor mechanism in the Euclidean signature with higher-derivative terms included (the Lorentzian version is discussed in \cite{Chen:2024gmc}). In particular, we would like to understand how the Euclidean solutions of \cite{Hegde:2023jmp,Chowdhury:2024ngg, Chen:2024gmc} fit in the framework discussed above.  Then, we can redo the analysis of \cite{Hegde:2024bmb} in  Euclidean supergravity. We hope to report on this in our future work.

\acknowledgments
We thank Aravind Aikot and, in particular, Subramanya Hegde for numerous helpful discussions and for carefully reading an earlier version of the manuscript. S.A. and A.B. thank the Chennai Mathematical Institute for its warm hospitality, during which much of this work was completed. The work of A.V. was partially supported by the SERB Core Research Grant CRG/2023/000545. The research of S.A. was supported in part by the National Research Foundation of Korea (NRF) grant funded by the Korean government (MSIT), Grant No. RS-2024-00449284; by the Sogang University Research Grant No. 202410008.01; and by the Basic Science Research Program of the NRF funded by the Ministry of Education through the Center for Quantum Spacetime (CQUeST), Grant No. RS-2020-NR049598.

\appendix
\section{Conventions}
\label{app:conventions}
\begin{equation}
    \{\gamma^a, \gamma^b\}=2\delta^{ab}\mathbb{I},
\end{equation}
\begin{equation}
    \gamma ^a\gamma^b=\delta^{ab}\mathbb{I}+\gamma^{ab},~~\gamma^{ab}=\frac{1}{2}(\gamma^a\gamma^b-\gamma^b\gamma^a),
\end{equation}
\begin{equation}
    \varepsilon^{0123}=1,~~\varepsilon_{0123}=1.
\end{equation}
\begin{align}
    \delta^{abcd}_{efgh}=4! \delta^a_{[e}\dots \delta^d_{h]}, ~~\varepsilon^{abcd}\varepsilon_{efgh}=\delta^{abcd}_{efgh},
\end{align}
\begin{equation}
    \gamma_5=\frac{1}{4!}\varepsilon_{abcd}\gamma^a\gamma^b\gamma^c\gamma^d, ~~\gamma_{abcd}=\varepsilon_{abcd}\gamma_5.
\end{equation}
\begin{equation}
    \{\gamma_{ab}, \gamma_c\}=2\gamma_{abc}= 2\gamma_{abcd}\gamma^d=2\varepsilon_{abcd}\gamma_5 \gamma^d, ~~ [\gamma_{ab}, \gamma_c]=4 \gamma_{[a}\delta_{b]c}.
\end{equation}
\begin{align}
    [\gamma_{ab}, \gamma^{cd}]=8\gamma_{[a}{}^{[d}\delta_{b]}{}^{c]},~~\{\gamma_{ab}, \gamma^{cd}\}=-4\delta_{[a}^{c}\delta_{b]}^{d}+2\varepsilon_{ab}{}^{cd}\gamma_{5}.
\end{align}
\begin{align}
T_{cd}^{\pm}\gamma^{cd}\gamma_{ab}\gamma_e \epsilon^i_{\pm}\,& =\Big(-4\delta ^{e}_{g}T_{ab}^{\pm}-8\delta_{g[a}T_{b]}{}^{e\pm}+8\delta^{e}_{[a}T_{b]g}^{\pm}\Big)\gamma^g \epsilon^i_{\pm}.
\end{align}

\bibliography{main}

@article{Sen:1995in,
    author = "Sen, Ashoke",
    title = "{Extremal black holes and elementary string states}",
    eprint = "hep-th/9504147",
    archivePrefix = "arXiv",
    reportNumber = "TIFR-TH-95-19",
    doi = "10.1142/S0217732395002234",
    journal = "Mod. Phys. Lett. A",
    volume = "10",
    pages = "2081--2094",
    year = "1995"
}

@article{Strominger:1996sh,
    author = "Strominger, Andrew and Vafa, Cumrun",
    title = "{Microscopic origin of the Bekenstein-Hawking entropy}",
    eprint = "hep-th/9601029",
    archivePrefix = "arXiv",
    reportNumber = "HUTP-96-A002, RU-96-01",
    doi = "10.1016/0370-2693(96)00345-0",
    journal = "Phys. Lett. B",
    volume = "379",
    pages = "99--104",
    year = "1996"
}

@article{Sen:2007qy,
    author = "Sen, Ashoke",
    title = "{Black Hole Entropy Function, Attractors and Precision Counting of Microstates}",
    eprint = "0708.1270",
    archivePrefix = "arXiv",
    primaryClass = "hep-th",
    doi = "10.1007/s10714-008-0626-4",
    journal = "Gen. Rel. Grav.",
    volume = "40",
    pages = "2249--2431",
    year = "2008"
}

@article{Cabo-Bizet:2018ehj,
    author = "Cabo-Bizet, Alejandro and Cassani, Davide and Martelli, Dario and Murthy, Sameer",
    title = "{Microscopic origin of the Bekenstein-Hawking entropy of supersymmetric AdS$_{5}$ black holes}",
    eprint = "1810.11442",
    archivePrefix = "arXiv",
    primaryClass = "hep-th",
    doi = "10.1007/JHEP10(2019)062",
    journal = "JHEP",
    volume = "10",
    pages = "062",
    year = "2019"
}

@article{Iliesiu:2021are,
    author = "Iliesiu, Luca V. and Kologlu, Murat and Turiaci, Gustavo J.",
    title = "{Supersymmetric indices factorize}",
    eprint = "2107.09062",
    archivePrefix = "arXiv",
    primaryClass = "hep-th",
    doi = "10.1007/JHEP05(2023)032",
    journal = "JHEP",
    volume = "05",
    pages = "032",
    year = "2023"
}

@article{Cassani:2019mms,
    author = "Cassani, Davide and Papini, Lorenzo",
    title = "{The BPS limit of rotating AdS black hole thermodynamics}",
    eprint = "1906.10148",
    archivePrefix = "arXiv",
    primaryClass = "hep-th",
    doi = "10.1007/JHEP09(2019)079",
    journal = "JHEP",
    volume = "09",
    pages = "079",
    year = "2019"
}

@article{Bobev:2020pjk,
    author = "Bobev, Nikolay and Charles, Anthony M. and Min, Vincent S.",
    title = "{Euclidean black saddles and AdS$_{4}$ black holes}",
    eprint = "2006.01148",
    archivePrefix = "arXiv",
    primaryClass = "hep-th",
    doi = "10.1007/JHEP10(2020)073",
    journal = "JHEP",
    volume = "10",
    pages = "073",
    year = "2020"
}

@article{Hristov:2022pmo,
    author = "Hristov, Kiril",
    title = "{The dark (BPS) side of thermodynamics in Minkowski$_{4}$}",
    eprint = "2207.12437",
    archivePrefix = "arXiv",
    primaryClass = "hep-th",
    doi = "10.1007/JHEP09(2022)204",
    journal = "JHEP",
    volume = "09",
    pages = "204",
    year = "2022"
}

@article{H:2023qko,
    author = "H., Anupam A. and Athira, P. V. and Chowdhury, Chandramouli and Sen, Ashoke",
    title = "{Logarithmic correction to BPS black hole entropy from supersymmetric index at finite temperature}",
    eprint = "2306.07322",
    archivePrefix = "arXiv",
    primaryClass = "hep-th",
    doi = "10.1007/JHEP03(2024)095",
    journal = "JHEP",
    volume = "03",
    pages = "095",
    year = "2024"
}

@article{Anupam:2023yns,
    author = "Anupam, A. H. and Chowdhury, Chandramouli and Sen, Ashoke",
    title = "{Revisiting logarithmic correction to five dimensional BPS black hole entropy}",
    eprint = "2308.00038",
    archivePrefix = "arXiv",
    primaryClass = "hep-th",
    doi = "10.1007/JHEP05(2024)070",
    journal = "JHEP",
    volume = "05",
    pages = "070",
    year = "2024"
}

@article{Boruch:2023gfn,
    author = "Boruch, Jan and Iliesiu, Luca V. and Murthy, Sameer and Turiaci, Gustavo J.",
    title = "{New forms of attraction: attractor saddles for the black hole index}",
    eprint = "2310.07763",
    archivePrefix = "arXiv",
    primaryClass = "hep-th",
    doi = "10.1007/JHEP04(2025)087",
    journal = "JHEP",
    volume = "04",
    pages = "087",
    year = "2025"
}

@article{Hegde:2023jmp,
    author = "Hegde, Subramanya and Virmani, Amitabh",
    title = "{Killing spinors for finite temperature Euclidean solutions at the BPS bound}",
    eprint = "2311.09427",
    archivePrefix = "arXiv",
    primaryClass = "hep-th",
    doi = "10.1007/JHEP02(2024)203",
    journal = "JHEP",
    volume = "02",
    pages = "203",
    year = "2024"
}

@article{Chowdhury:2024ngg,
    author = "Chowdhury, Chandramouli and Sen, Ashoke and Shanmugapriya, P. and Virmani, Amitabh",
    title = "{Supersymmetric index for small black holes}",
    eprint = "2401.13730",
    archivePrefix = "arXiv",
    primaryClass = "hep-th",
    doi = "10.1007/JHEP04(2024)136",
    journal = "JHEP",
    volume = "04",
    pages = "136",
    year = "2024"
}

@article{Chen:2024gmc,
    author = "Chen, Yiming and Murthy, Sameer and Turiaci, Gustavo J.",
    title = "{Gravitational index of the heterotic string}",
    eprint = "2402.03297",
    archivePrefix = "arXiv",
    primaryClass = "hep-th",
    doi = "10.1007/JHEP09(2024)041",
    journal = "JHEP",
    volume = "09",
    pages = "041",
    year = "2024"
}

@article{Cassani:2024kjn,
    author = "Cassani, Davide and Ruip{\'e}rez, Alejandro and Turetta, Enrico",
    title = "{Localization of the 5D supergravity action and Euclidean saddles for the black hole index}",
    eprint = "2409.01332",
    archivePrefix = "arXiv",
    primaryClass = "hep-th",
    doi = "10.1007/JHEP12(2024)086",
    journal = "JHEP",
    volume = "12",
    pages = "086",
    year = "2024"
}

@article{Hegde:2024bmb,
    author = "Hegde, Subramanya and Sen, Ashoke and Shanmugapriya, P. and Virmani, Amitabh",
    title = "{Supersymmetric index for half BPS black holes in N=2 supergravity with higher curvature corrections}",
    eprint = "2411.08260",
    archivePrefix = "arXiv",
    primaryClass = "hep-th",
    doi = "10.1007/JHEP02(2025)131",
    journal = "JHEP",
    volume = "02",
    pages = "131",
    year = "2025"
}

@article{Adhikari:2024zif,
    author = "Adhikari, Soumya and Dharanipragada, Pavan and Goswami, Kaberi and Virmani, Amitabh",
    title = "{Attractor saddle for 5D black hole index}",
    eprint = "2411.12413",
    archivePrefix = "arXiv",
    primaryClass = "hep-th",
    doi = "10.1007/JHEP03(2025)180",
    journal = "JHEP",
    volume = "03",
    pages = "180",
    year = "2025"
}

@article{Boruch:2025qdq,
    author = "Boruch, Jan and Emparan, Roberto and Iliesiu, Luca V. and Murthy, Sameer",
    title = "{The gravitational index of 5d black holes and black strings}",
    eprint = "2501.17909",
    archivePrefix = "arXiv",
    primaryClass = "hep-th",
    doi = "10.1007/JHEP06(2025)145",
    journal = "JHEP",
    volume = "06",
    pages = "145",
    year = "2025"
}

@article{Bandyopadhyay:2025jbc,
    author = "Bandyopadhyay, Subhodip and Punia, Gurmeet Singh and Srivastava, Yogesh K. and Virmani, Amitabh",
    title = "{The gravitational index of a small black ring}",
    eprint = "2504.09982",
    archivePrefix = "arXiv",
    primaryClass = "hep-th",
    doi = "10.1007/JHEP07(2025)200",
    journal = "JHEP",
    volume = "07",
    pages = "200",
    year = "2025"
}

@article{Boruch:2025biv,
    author = "Boruch, Jan and Iliesiu, Luca V. and Murthy, Sameer and Turiaci, Gustavo J.",
    title = "{Multicentered black hole saddles for supersymmetric indices}",
    eprint = "2507.07166",
    archivePrefix = "arXiv",
    primaryClass = "hep-th",
    month = "7",
    year = "2025"
}

@article{Cassani:2025iix,
    author = "Cassani, Davide and Ruip{\'e}rez, Alejandro and Turetta, Enrico",
    title = "{Bubbling saddles of the gravitational index}",
    eprint = "2507.12650",
    archivePrefix = "arXiv",
    primaryClass = "hep-th",
    doi = "10.21468/SciPostPhys.19.5.134",
    journal = "SciPost Phys.",
    volume = "19",
    pages = "134",
    year = "2025"
}

@article{Boruch:2025sie,
    author = "Boruch, Jan and Emparan, Roberto and Iliesiu, Luca V. and Murthy, Sameer",
    title = "{Novel black saddles for 5d gravitational indices and the index enigma}",
    eprint = "2510.23699",
    archivePrefix = "arXiv",
    primaryClass = "hep-th",
    month = "10",
    year = "2025"
}

@inbook{Cassani:2025sim,
    author = "Cassani, Davide and Murthy, Sameer",
    title = "{Quantum black holes: supersymmetry and exact results}",
    eprint = "2502.15360",
    archivePrefix = "arXiv",
    primaryClass = "hep-th",
    month = "2",
    year = "2025"
}

@article{Dabholkar:2010uh,
    author = "Dabholkar, Atish and Gomes, Joao and Murthy, Sameer",
    title = "{Quantum black holes, localization and the topological string}",
    eprint = "1012.0265",
    archivePrefix = "arXiv",
    primaryClass = "hep-th",
    doi = "10.1007/JHEP06(2011)019",
    journal = "JHEP",
    volume = "06",
    pages = "019",
    year = "2011"
}

@article{Sen:2008vm,
    author = "Sen, Ashoke",
    title = "{Quantum Entropy Function from AdS(2)/CFT(1) Correspondence}",
    eprint = "0809.3304",
    archivePrefix = "arXiv",
    primaryClass = "hep-th",
    doi = "10.1142/S0217751X09045893",
    journal = "Int. J. Mod. Phys. A",
    volume = "24",
    pages = "4225--4244",
    year = "2009"
}

@article{LopesCardoso:2000qm,
    author = "Lopes Cardoso, Gabriel and de Wit, Bernard and Kappeli, Jurg and Mohaupt, Thomas",
    title = "{Stationary BPS solutions in N=2 supergravity with R**2 interactions}",
    eprint = "hep-th/0009234",
    archivePrefix = "arXiv",
    reportNumber = "SPIN-2000-21, SPIN-00-21, ITP-UU-00-24, SU-ITP-00-19",
    doi = "10.1088/1126-6708/2000/12/019",
    journal = "JHEP",
    volume = "12",
    pages = "019",
    year = "2000"
}

@article{deWit:2017cle,
    author = "de Wit, Bernard and Reys, Valentin",
    title = "{Euclidean supergravity}",
    eprint = "1706.04973",
    archivePrefix = "arXiv",
    primaryClass = "hep-th",
    doi = "10.1007/JHEP12(2017)011",
    journal = "JHEP",
    volume = "12",
    pages = "011",
    year = "2017"
}

@article{Cortes:2003zd,
    author = "Cortes, Vicente and Mayer, Christoph and Mohaupt, Thomas and Saueressig, Frank",
    title = "{Special geometry of Euclidean supersymmetry. 1. Vector multiplets}",
    eprint = "hep-th/0312001",
    archivePrefix = "arXiv",
    reportNumber = "FSU-TPI-12-03",
    doi = "10.1088/1126-6708/2004/03/028",
    journal = "JHEP",
    volume = "03",
    pages = "028",
    year = "2004"
}

@article{Cortes:2005uq,
    author = "Cortes, Vicente and Mayer, Christoph and Mohaupt, Thomas and Saueressig, Frank",
    title = "{Special geometry of euclidean supersymmetry. II. Hypermultiplets and the c-map}",
    eprint = "hep-th/0503094",
    archivePrefix = "arXiv",
    reportNumber = "ITP-05-06, SPIN-05-04, FSU-TPI-02-05",
    doi = "10.1088/1126-6708/2005/06/025",
    journal = "JHEP",
    volume = "06",
    pages = "025",
    year = "2005"
}

@article{Cortes:2009cs,
    author = "Cortes, Vicente and Mohaupt, Thomas",
    title = "{Special Geometry of Euclidean Supersymmetry III: The Local r-map, instantons and black holes}",
    eprint = "0905.2844",
    archivePrefix = "arXiv",
    primaryClass = "hep-th",
    reportNumber = "LTH-830",
    doi = "10.1088/1126-6708/2009/07/066",
    journal = "JHEP",
    volume = "07",
    pages = "066",
    year = "2009"
}

@article{Jeon:2018kec,
    author = "Jeon, Imtak and Murthy, Sameer",
    title = "{Twisting and localization in supergravity: equivariant cohomology of BPS black holes}",
    eprint = "1806.04479",
    archivePrefix = "arXiv",
    primaryClass = "hep-th",
    doi = "10.1007/JHEP03(2019)140",
    journal = "JHEP",
    volume = "03",
    pages = "140",
    year = "2019"
}

@article{Ciceri:2023mjl,
    author = "Ciceri, Axel and Jeon, Imtak and Murthy, Sameer",
    title = "{Localization on AdS$_{3}${\texttimes} S$^{2}$. Part I. The 4d/5d connection in off-shell Euclidean supergravity}",
    eprint = "2301.08084",
    archivePrefix = "arXiv",
    primaryClass = "hep-th",
    doi = "10.1007/JHEP07(2023)218",
    journal = "JHEP",
    volume = "07",
    pages = "218",
    year = "2023"
}

@article{LopesCardoso:1998tkj,
    author = "Lopes Cardoso, Gabriel and de Wit, Bernard and Mohaupt, Thomas",
    title = "{Corrections to macroscopic supersymmetric black hole entropy}",
    eprint = "hep-th/9812082",
    archivePrefix = "arXiv",
    reportNumber = "SPIN-1998-11, THU-98-44",
    doi = "10.1016/S0370-2693(99)00227-0",
    journal = "Phys. Lett. B",
    volume = "451",
    pages = "309--316",
    year = "1999"
}

@article{Mohaupt:2000mj,
    author = "Mohaupt, Thomas",
    title = "{Black hole entropy, special geometry and strings}",
    eprint = "hep-th/0007195",
    archivePrefix = "arXiv",
    doi = "10.1002/1521-3978(200102)49:1/3<3::AID-PROP3>3.0.CO;2-#",
    journal = "Fortsch. Phys.",
    volume = "49",
    pages = "3--161",
    year = "2001"
}

@article{Bhattacharjee:2025qro,
    author = "Bhattacharjee, Abhinava and Hegde, Subramanya and Sahoo, Bindusar",
    title = "{Only Flat Space is Full BPS in Four Dimensional N=3 and N=4 Supergravity}",
    eprint = "2505.00638",
    archivePrefix = "arXiv",
    primaryClass = "hep-th",
    month = "5",
    year = "2025"
}

@article{Ferrara:1977mv,
    author = "Ferrara, S. and Zumino, B.",
    title = "{Structure of Conformal Supergravity}",
    reportNumber = "CERN-TH-2418",
    doi = "10.1016/0550-3213(78)90548-5",
    journal = "Nucl. Phys. B",
    volume = "134",
    pages = "301--326",
    year = "1978"
}

@article{deWit:1979dzm,
    author = "de Wit, B. and van Holten, J. W. and Van Proeyen, Antoine",
    title = "{Transformation Rules of N=2 Supergravity Multiplets}",
    reportNumber = "KUL-TF-79/034",
    doi = "10.1016/0550-3213(80)90125-X",
    journal = "Nucl. Phys. B",
    volume = "167",
    pages = "186",
    year = "1980"
}

@article{Bergshoeff:1980is,
    author = "Bergshoeff, E. and de Roo, M. and de Wit, B.",
    title = "{Extended Conformal Supergravity}",
    reportNumber = "NIKHEF-H/80-07",
    doi = "10.1016/0550-3213(81)90465-X",
    journal = "Nucl. Phys. B",
    volume = "182",
    pages = "173--204",
    year = "1981"
}

@article{Bergshoeff:1985mz,
    author = "Bergshoeff, E. and Sezgin, E. and Van Proeyen, Antoine",
    editor = "Salam, A. and Sezgin, E.",
    title = "{Superconformal Tensor Calculus and Matter Couplings in Six-dimensions}",
    reportNumber = "IC-85-112",
    doi = "10.1016/0550-3213(86)90503-1",
    journal = "Nucl. Phys. B",
    volume = "264",
    pages = "653",
    year = "1986",
    note = "[Erratum: Nucl.Phys.B 598, 667 (2001)]"
}

@article{Fujita:2001kv,
    author = "Fujita, Tomoyuki and Ohashi, Keisuke",
    title = "{Superconformal tensor calculus in five-dimensions}",
    eprint = "hep-th/0104130",
    archivePrefix = "arXiv",
    reportNumber = "KUNS-1716",
    doi = "10.1143/PTP.106.221",
    journal = "Prog. Theor. Phys.",
    volume = "106",
    pages = "221--247",
    year = "2001"
}

@article{Butter:2013goa,
    author = "Butter, Daniel and Kuzenko, Sergei M. and Novak, Joseph and Tartaglino-Mazzucchelli, Gabriele",
    title = "{Conformal supergravity in three dimensions: New off-shell formulation}",
    eprint = "1305.3132",
    archivePrefix = "arXiv",
    primaryClass = "hep-th",
    reportNumber = "NIKHEF-2013-014",
    doi = "10.1007/JHEP09(2013)072",
    journal = "JHEP",
    volume = "09",
    pages = "072",
    year = "2013"
}

@article{Butter:2013rba,
    author = "Butter, Daniel and Kuzenko, Sergei M. and Novak, Joseph and Tartaglino-Mazzucchelli, Gabriele",
    title = "{Conformal supergravity in three dimensions: Off-shell actions}",
    eprint = "1306.1205",
    archivePrefix = "arXiv",
    primaryClass = "hep-th",
    reportNumber = "NIKHEF-2013-018",
    doi = "10.1007/JHEP10(2013)073",
    journal = "JHEP",
    volume = "10",
    pages = "073",
    year = "2013"
}

@article{Behrndt:1996jn,
    author = "Behrndt, Klaus and Lopes Cardoso, Gabriel and de Wit, Bernard and Kallosh, Renata and Lust, Dieter and Mohaupt, Thomas",
    title = "{Classical and quantum N=2 supersymmetric black holes}",
    eprint = "hep-th/9610105",
    archivePrefix = "arXiv",
    reportNumber = "CERN-TH-96-276, HUB-EP-96-53, SU-ITP-96-41, THU-96-34",
    doi = "10.1016/S0550-3213(97)00028-X",
    journal = "Nucl. Phys. B",
    volume = "488",
    pages = "236--260",
    year = "1997"
}
\bibliographystyle{JHEP}

\end{document}